\documentclass[structabstract]{aa}  

\usepackage{graphicx}
\usepackage{natbib}
\usepackage{bm}
\usepackage{txfonts}
\begin{document}
   \title{The subcritical baroclinic instability in local accretion disc models}

   \author{Geoffroy Lesur\inst{1} and John C. B. Papaloizou \inst{1}         }
   \institute{Department of Applied Mathematics and Theoretical Physics, University of Cambridge, Centre for Mathematical Sciences,
Wilberforce Road, Cambridge CB3 0WA, UK \\
              \email{g.lesur@damtp.cam.ac.uk}
                     }

   \date{Received date / Accepted 13 January 2010}

 
  \abstract
   {The presence of vortices in accretion discs has been debated for more than a decade. Baroclinic instabilities might be a way to generate these vortices in the presence of a radial entropy gradient. However, the nature of these instabilities is still unclear and 3D parametric instabilities can lead to the rapid destruction of these vortices.}
   {We present new results exhibiting a subcritical baroclinic instability (SBI) in local shearing box models. We describe the 2D and 3D behaviour of this instability using numerical simulations and we present a simple analytical model describing the underlying physical process. }
   {We investigate the SBI in local shearing boxes, using either the incompressible Boussinesq approximation or a fully compressible model. We explore the parameter space varying several local dimensionless parameters and we isolate the regime relevant for the SBI. 3D shearing boxes are also investigated using high resolution spectral methods to resolve both the SBI and 3D parametric instabilities. }
   {A subcritical baroclinic instability is observed in flows stable for the Solberg-Ho\"iland criterion using local simulations. This instability is found to be a nonlinear (or subcritical) instability, which cannot be described by ordinary linear approaches. It requires a radial entropy gradient weakly unstable for the Schwartzchild criterion and a strong thermal diffusivity (or equivalently a short cooling time). In compressible simulations, the instability produces density waves which transport angular momentum outward with typically $\alpha\lesssim 3\times 10^{-3}$, the exact value depending on the background temperature profile. Finally, the instability survives in 3D, vortex cores becoming turbulent due to parametric instabilities. }
   {The subcritical baroclinic instability is a robust phenomenon, which can be captured using local simulations. The instability survives in 3D thanks to a balance between the 2D SBI and 3D parametric instabilities. Finally, this instability can lead to a \emph{weak} outward transport of angular momentum, due to the generation of density waves by the vortices.}


   \maketitle
%

\section{Introduction}
The existence of long-lived vortices in accretion discs was first proposed by \cite{W44} in a model of planet formation. This idea was reintroduced by \cite{BS95} to accelerate the formation of planetesimals through a dust trapping process. Moreover, large scale vortices can lead to a significant transport of angular momentum thanks to the production of density waves \citep{JG05b}.
Many physical processes have been introduced in the literature to justify the existence of these vortices such as Rossby wave instabilities \citep{LLC99}, planetary gap instabilities \citep{VE06,VAAP07}, 3D circulation models \citep{BM05}, MHD turbulence \citep{FN05} and the global baroclinic instability \citep{KB03}. 

Baroclinic instabilities in accretion discs has regained interest in the past few years. A first version of these instabilities (the global baroclinic instability or GBI) was mentioned by \cite{KB03} using a purely numerical approach and considering a disc with a radial entropy gradient. However, the presence of this instability was affected by changing the numerical scheme used in the simulations, casting strong doubts on this instability as a real physical processes. The linear properties of the GBI were then investigated by \cite{K04} and \cite{JG05}. However, only transient growths were found in this context. \cite{K04} speculated that these transient growths could lead to a nonlinear instability explaining the result of \cite{KB03}, but it is still unclear whether transient growths are relevant for nonlinear instabilities \citep[see e.g.][and reference therein for a complete discussion of this point]{LL05}. The nonlinear problem was then studied in local shearing boxes by \cite{JG06} using fully compressible numerical methods. These authors found no instability in the Keplerian disc regime and concluded that the GBI was ``either global or nonexistent''. 

Nevertheless, baroclinic processes were studied again by \cite{PJS07} using anelastic global simulations and including new physical processes. As in \cite{KB03}, a weak radial entropy gradient was imposed in the simulation. However, a cooling function was also included in their model, in order to force the system to relax to the imposed radial temperature profile. In these simulations, \cite{PJS07} observed spontaneous formation of vortices with radial entropy gradients compatible with accretion disc thermodynamics. In a subsequent paper \citep{PSJ07}, these vortices were found to survive for several hundreds orbits. According to the authors, their disagreement with \cite{JG06} was due to a larger Reynolds number and the use of spectral methods, a possibility already mentioned by \cite{JG06}.

In this paper, we revisit baroclinic instabilities in a local setup (namely the shearing box) both in the incompressible-Boussinesq approximation and in fully compressible simulations. The aim of this paper is to clarify several of the points which have been discussed in the literature for the past 6 years about the existence, the nature and the properties of baroclinic instabilities. We show that a subcritical baroclinic instability (or shortly SBI) \emph{does exist} in shearing boxes. This instability is \emph{nonlinear} (or subcritical) and \emph{strongly linked to thermal diffusion}, a point already mentioned by \cite{PSJ07}. 
We also present results concerning the behaviour of the SBI in 3 dimensions, as vortices are known to be unstable because of parametric instabilities \citep{LP09}. We begin in \S\ref{equations} by presenting the equations, introducing the dimensionless numbers and describing the numerical methods used in the rest of the paper. \S\ref{2Dincomp} is dedicated to 2D incompressible-Boussinesq simulations and a qualitative understanding of the instability. We present in \S\ref{compressibility} our results in compressible shearing boxes and in \S\ref{3Dincomp} the 3D behaviour of the instability. Conclusions and implications of our findings are discussed in \S\ref{conclusions}.

\section{Local model\label{equations}}
\subsection{Equations}
In the following, we will assume a local model for the accretion disc, following the shearing sheet approximation. The reader may consult \cite{HGB95}, \cite{B03} and \cite{RU08} for an extensive discussion of the properties and limitations of this model. As a simplification, we will assume the flow is incompressible, consistently with the small shearing box model \citep{RU08}. The shearing-box equations are found by considering a Cartesian box centred at $r=R_0$, rotating with the disc at angular velocity $\Omega=\Omega(R_0)$. We finally introduce in this box a radial stratification using the Boussinesq approximation \citep{SV60}. 
Defining $r-R_0 \rightarrow x$ and $R_0\phi \rightarrow y$ as in \cite{HGB95}, one eventually obtains the following set of equations:
\begin{eqnarray}
\nonumber  \partial_t \bm{u}+\bm{u\cdot\nabla} \bm{u}&=&-\bm{\nabla} \Pi
-2\bm{\Omega \times u}\\
\label{motiongeneral}& &2\Omega S x \bm{e_x}-\Lambda N^2\theta \bm{e_x}+\nu\Delta\bm{u},\\
\label{entropygeneral}\partial_t \theta +\bm{u\cdot\nabla}\theta&=&u_x/\Lambda+\mu\Delta\theta\\
\label{divv} \bm{\nabla \cdot u}&=&0,
\end{eqnarray}
where $\bm{u}=u_x\bm{e_x}+u_y\bm{e_y}+u_z\bm{e_z}$ is the fluid velocity, $\theta$ the potential temperature deviation, $\nu$ the kinematic viscosity and $\mu$ the thermal diffusivity.
In these equations, we have defined the mean shear $S=-r\partial_r \Omega$, which is set to $S=(3/2)\Omega$ for a Keplerian disc. One can check easily that the velocity field $\bm{U}=-Sx\bm{e_y}$ is a steady solution of these equations. In the following we will consider the evolutions of the perturbations $\bm{v}$ (not necessarily small) of this profile defined by $\bm{v}=\bm{u}-\bm{U}$.

The generalised pressure $\Pi$ is calculated solving a Poisson equation with the incompressibility condition (\ref{divv}). For homogeneity and consistency with the traditional Boussinesq approach, we have introduced a stratification length $\Lambda$.  Note however that $\Lambda$ disappears from the dynamical properties of these equations as one can renormalise the variables defining $\theta'\equiv\Lambda \theta$.
The stratification itself is controlled by the Brunt-V\"ais\"al\"a frequency $N$, defined for a perfect gas by
\begin{equation}
N^2=-\frac{1}{\gamma \rho}\frac{\partial P}{\partial R}\frac{\partial}{\partial R}\ln\Big(\frac{P}{\rho^\gamma}\Big),
\end{equation}
where $P$ and $\rho$ are assumed to be the background equilibrium profile and $\gamma$ is the adiabatic index. With these notations, one can recover the ordinary thermodynamical variables as $\theta\equiv \delta \rho /\rho$, where $\delta \rho$ is the density perturbation and $\rho$ is the background density. The stratification length is then defined by $\Lambda\equiv -\partial_R P /\rho N^2$.

\subsection{Dimensionless numbers}
The system described above involves several physical processes. To clarify the regime in which we are working and to make easier comparisons with previous work, we define the following dimensionless numbers:
\begin{itemize}
\item The Richardson number $Ri=N^2/S^2$ compares the shear timescale to the buoyancy timescale. This definition is equivalent to the definition of \cite{JG05}.
\item The Peclet number $Pe=SL^2/\mu$.
\item The Reynolds number $Re=SL^2/\nu$.
\end{itemize}
$L$ is a typical scale of the system, chosen to be the radial box size ($L_x$) in our notations.

The linear stability properties of this flow are quite well understood. The flow is linearly stable for axisymmetric perturbations when the Solberg-Ho\"iland criterion is satisfied. This criterion may be written locally as:
\begin{equation}
\label{solberg}2\Omega(2\Omega -S)+N^2>0 \quad \quad \mathrm{Stability,}
\end{equation}
or equivalently in a Keplerian disc, $Ri>-4/9$. Another linear stability criterion, the Schwarzschild criterion, is often used for convectively stable flows \emph{without rotation nor shear}. This criterion reads
\begin{equation}
\label{schwarzschild} N^2>0 \quad \quad \mathrm{Stability},
\end{equation}
or equivalently $Ri>0$. As we will see in the following, this criterion is the relevant one for the SBI.

When viscosity and thermal diffusivity are included, the Solberg-Ho\"iland criterion is modified, potentially leading to the Goldreich-Schubert-Fricke (GSF) instability \citep{GS67,F68}. The stability criterion is in that case
\begin{equation}
\mu 2\Omega(2\Omega -S)+\nu N^2>0 \quad \quad \mathrm{Stability.}
\end{equation}
This criterion is satisfied when both (\ref{solberg}) is verified and $\nu/\mu<1$, which corresponds to the regime studied in this paper. 

\subsection{Numerical methods}\label{Numerics}
We have used two different codes to study this instability. When using the Boussineq model (\S\ref{2Dincomp} and \S\ref{3Dincomp}), we have used SNOOPY, a 3D incompressible spectral code. This code uses an implicit scheme for thermal and viscous diffusion, allowing us to study simulations with small $Pe$ without any strong constrain on the CFL condition. The spectral algorithm of SNOOPY has now been used in several hydrodynamic and magnetohydrodynamic studies \citep[e.g.][]{LL05,LL07} and is freely available on the author's website. For compressible simulations (\S\ref{compressibility}), we have used NIRVANA \citep{ZY97}. NIRVANA  has been used frequently in the past to study
various problems involving MHD turbulence in the shearing box \citep{FP06,PNS04}.

In the following, we use the shearing sheet boundary conditions \citep{HGB95} in the radial direction. This is made possible by the use of the Boussinesq approximation in which only the \emph{gradients} of the background profile appear explicitly, through $\Lambda$ and $N$. Therefore, when $\Lambda$ and $N$ are constant through the box, one can assume that the thermodynamic fluctuation $\theta$ is shear-periodic, consistently with the shearing-sheet approximation.

\section{2D subcritical baroclinic instability in incompressible flows\label{2Dincomp}}
To start our investigation, we consider the simplest case of a 2D $(x,y)$ problem in an infinitely thin disc. This setup is the local equivalent of the 2D global anelastic setup of \cite{PJS07}, and one expects to find similar properties in both cases if the instability is local.
In two dimensions, it is easier to consider the vorticity equation instead of (\ref{motiongeneral}). Defining the vertical vorticity of the pertubations by $\bm{\omega}=\partial_x v_y-\partial_y v_x$, we have:
\begin{equation}
\partial_t\omega+\bm{v\cdot\nabla}\omega-Sx\partial_y\omega=\Lambda N^2\partial_y \theta +\nu\Delta \omega
\end{equation}
In this formulation, the only source of enstrophy $\langle \omega^2/2 \rangle=\int dxdy \omega^2/2$ is the baroclinic term $\Lambda N^2\partial_y \theta$ which will be shown to play an important role for the instability.

The simulations presented in this section are computed in a square domain $L_x=L_y=1$ with a resolution of $512\times 512$ grid points using our spectral code. When not explicitly mentioned, the time unit is the shear timescale $S^{-1}$. We choose the fiducial parameters to be $Re=4\times10^5$ ; $Pe=4\times10^3$ and $Ri=-0.01$, in order to have a flow linearly stable for the Solberg-Ho\"iland criterion (\ref{solberg}). These parameters are close to the one used by \cite{PJS07} and are compatible with disc thermodynamics.

\subsection{Influence of initial perturbations}
\begin{figure}
   \centering
   \includegraphics[width=1.0\linewidth]{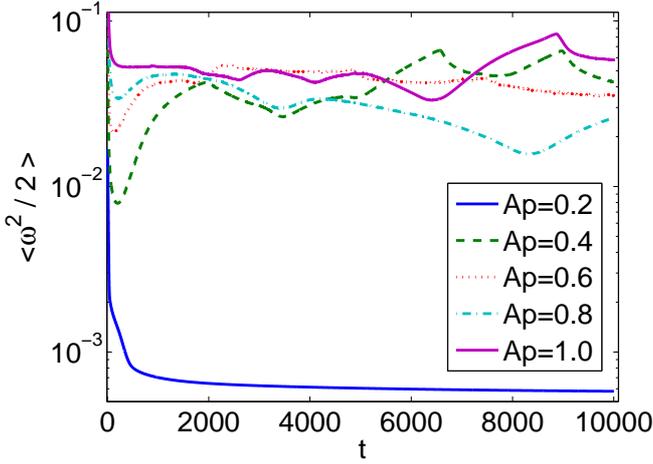}
   \caption{Volume averaged enstrophy for several initial amplitude perturbation ($Ap$) in arbitrary units. A subcritical instability is observed for finite amplitude perturbations between $Ap=0.2$ and $Ap=0.4$.}
              \label{subcrit}%
\end{figure}

In the first numerical experiment, we choose to test the effect of the initial conditions, keeping constant the dimensionless parameters. In our initial conditions, we excite randomly the largest wavelengths of the vorticity field with an amplitude $Ap$. This initial condition is slightly different from the one used by \cite{PJS07,PSJ07} who introduced perturbations in the temperature field.
In each experiment we modify the amplitude of the initial perturbation $Ap$, and follow the time evolution of the total enstrophy (Fig.~\ref{subcrit})

The numerical results indicates clearly the presence of a \emph{nonlinear} or \emph{subcritical} transition in the flow. Indeed, we find the ``instability'' for large enough initial perturbations. This also confirms the result of \cite{JG06}: there is no linear instability in the presence of a weak radial stratification. Moreover, one of the reasons that \cite{JG06} did not observe any transition could be because of the weak initial perturbations they used (between $10^{-12}$ and $10^{-4}$). Using similar initial conditions, we did not observe any transition either. The existence of the instability in our simulations was checked by doubling the resolution ($1024\times 1024$), keeping constant all the dimensionless numbers. We found no significant difference between high and low resolution results, showing that our simulations were converged for this problem.

\begin{figure*}
   \centering
   \includegraphics[width=0.33\linewidth]{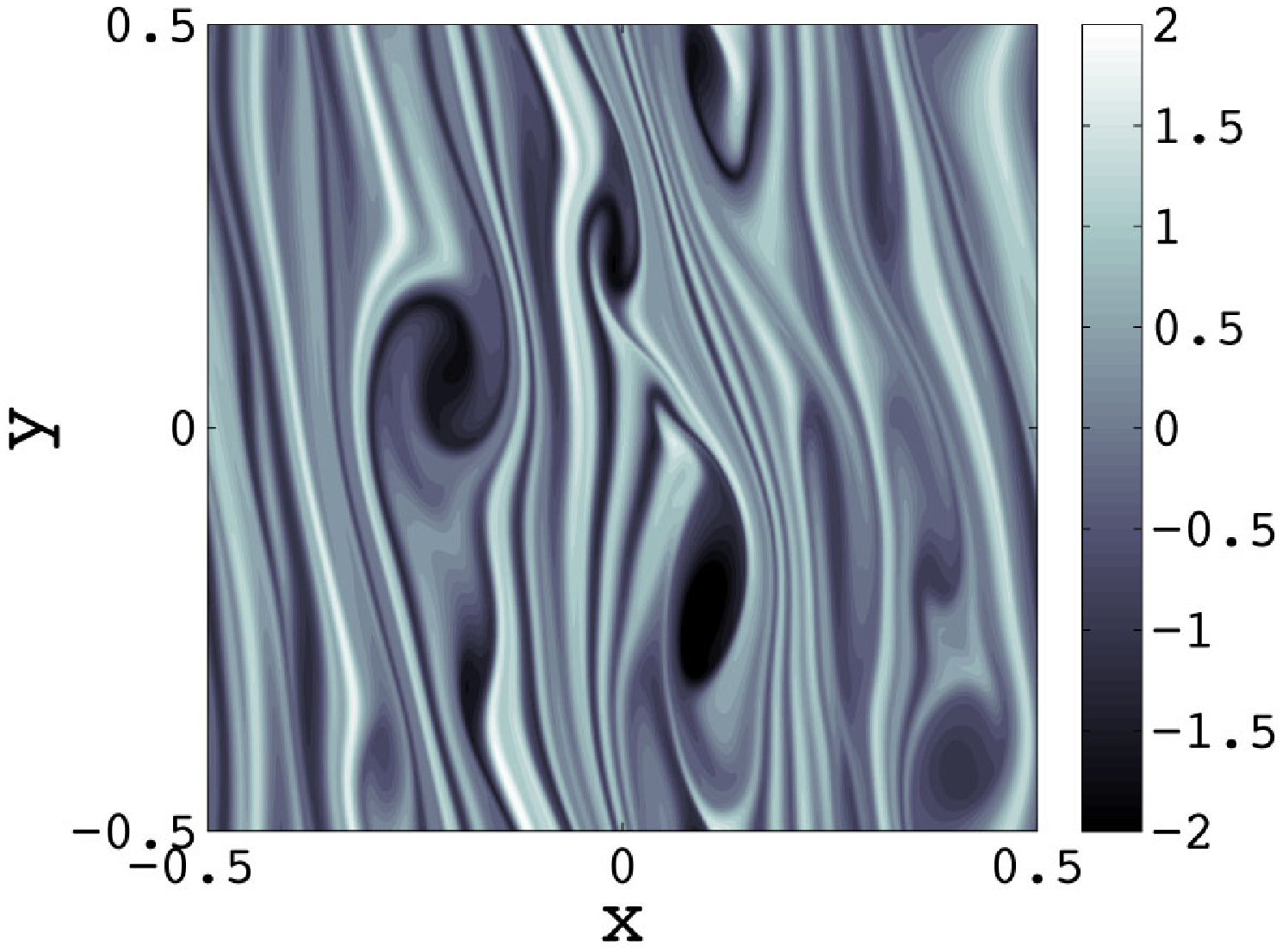}
   \includegraphics[width=0.33\linewidth]{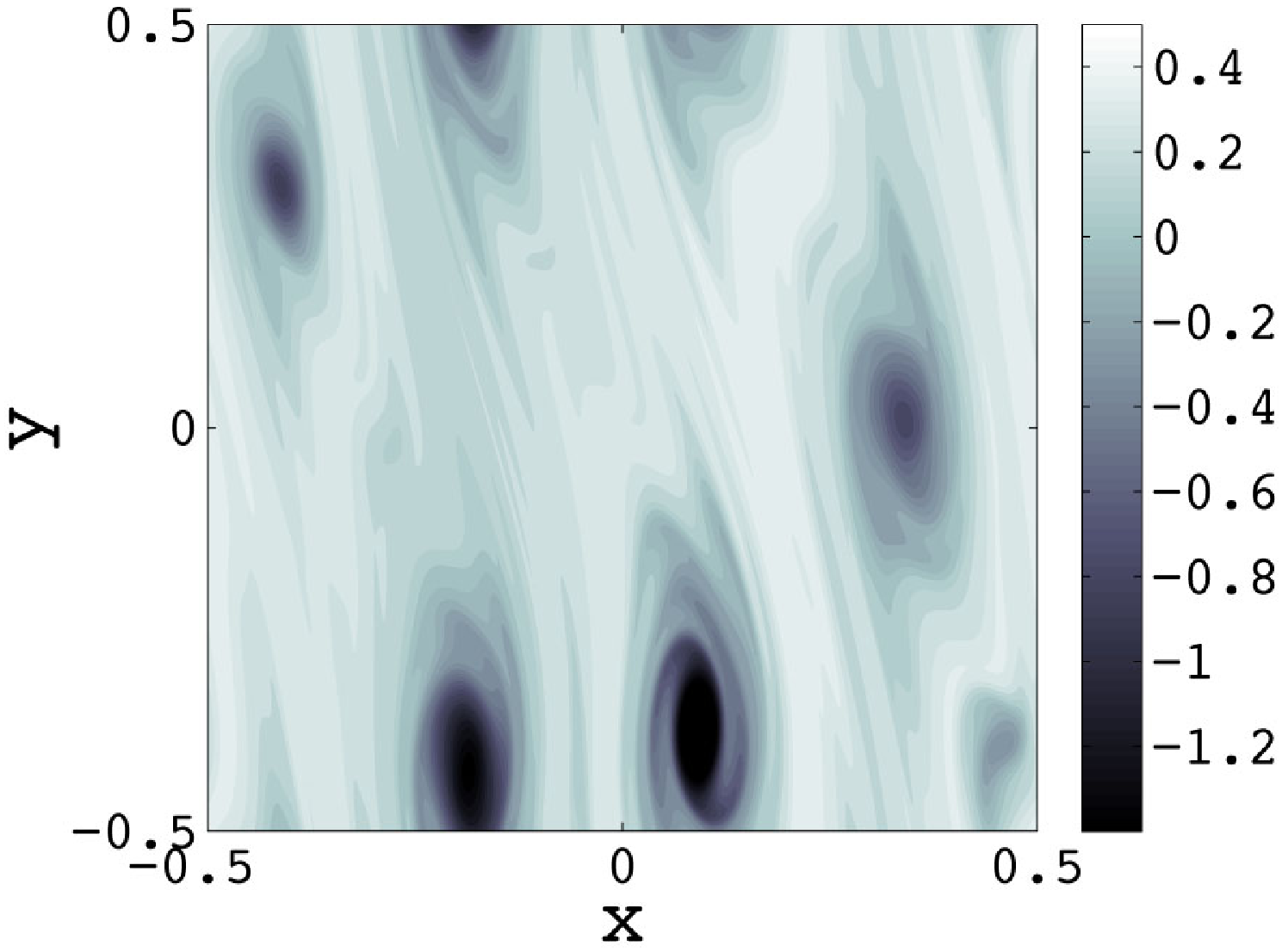}
   \includegraphics[width=0.33\linewidth]{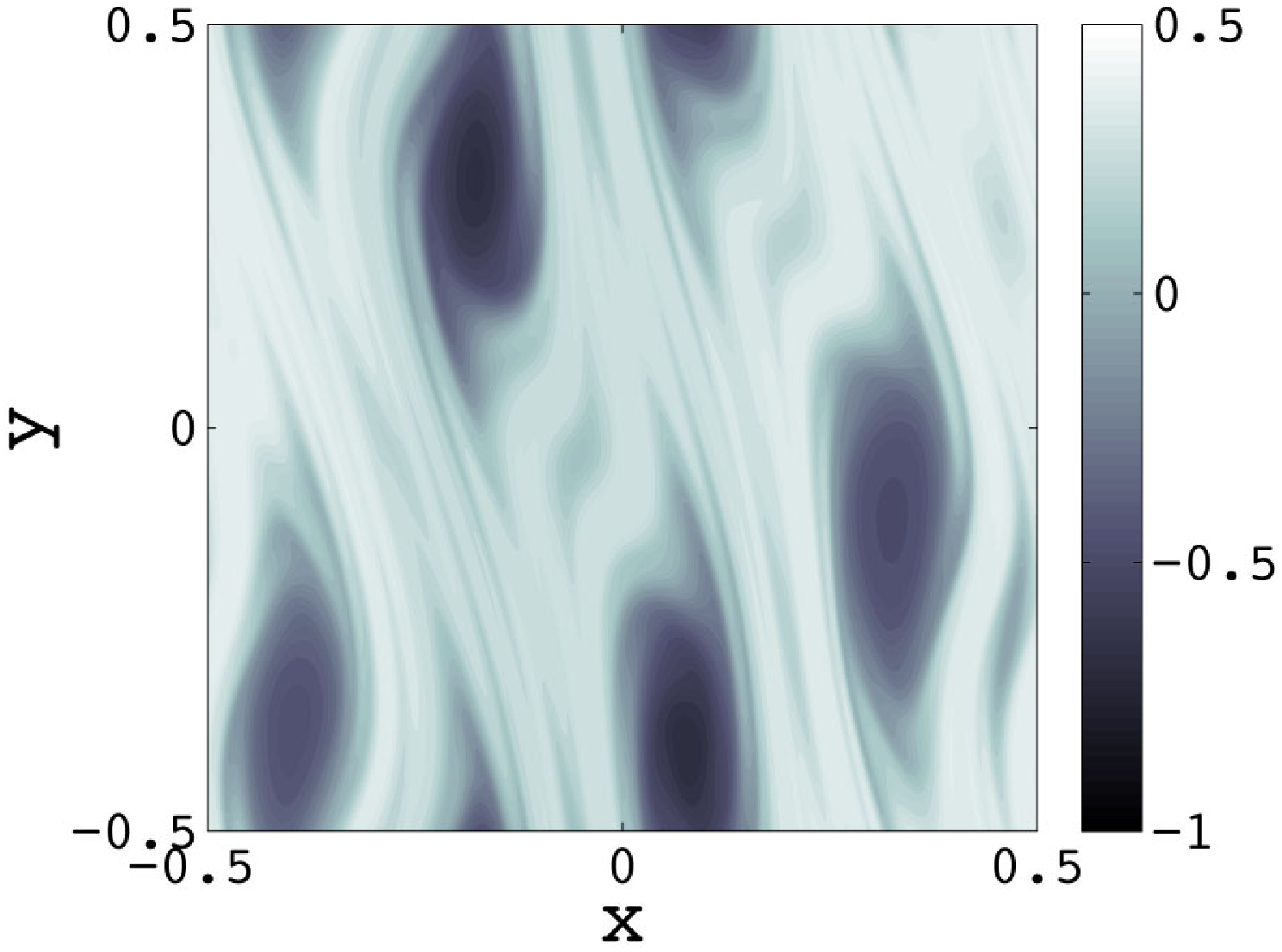}
   \includegraphics[width=0.33\linewidth]{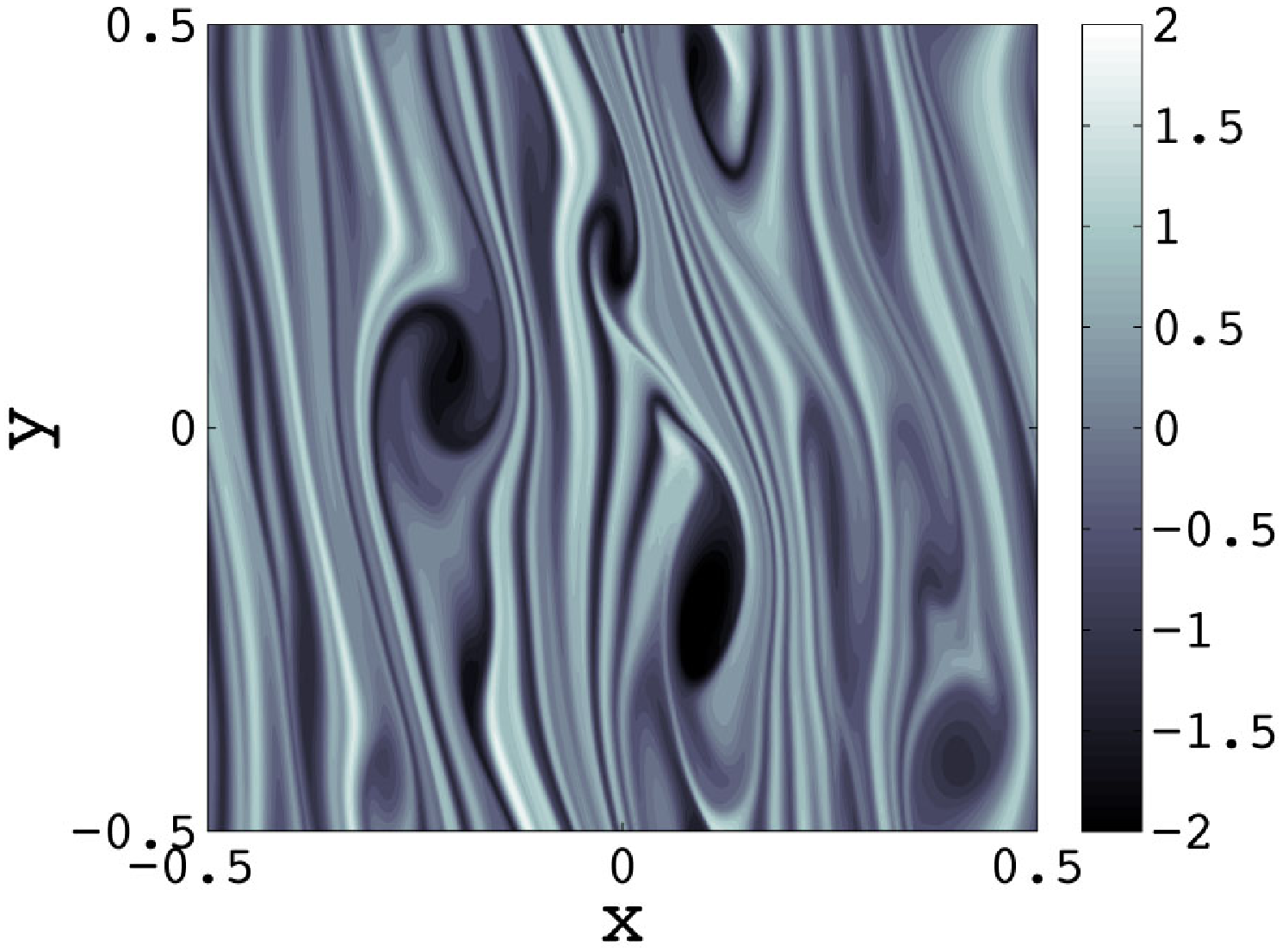}
   \includegraphics[width=0.33\linewidth]{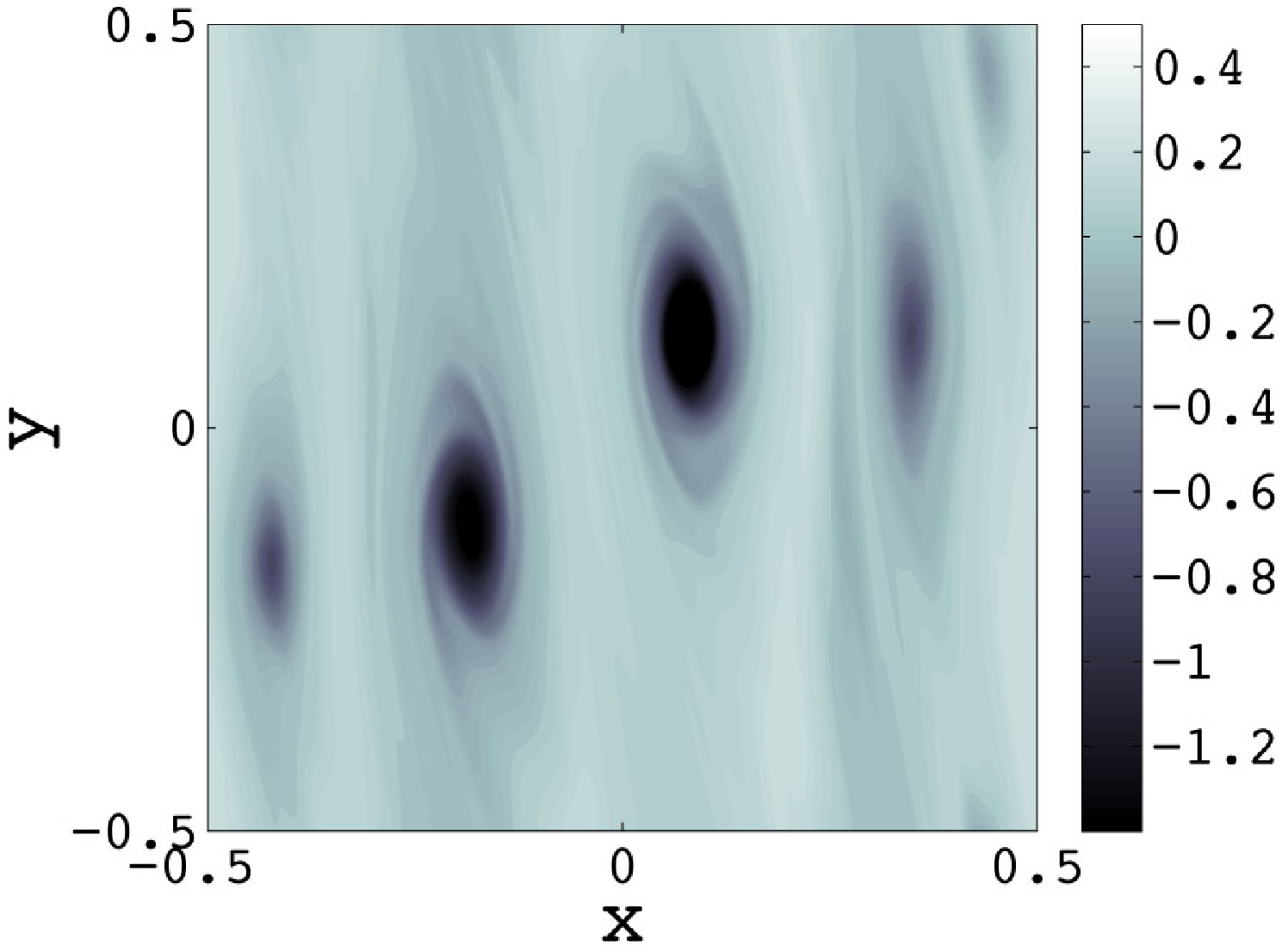}
   \includegraphics[width=0.33\linewidth]{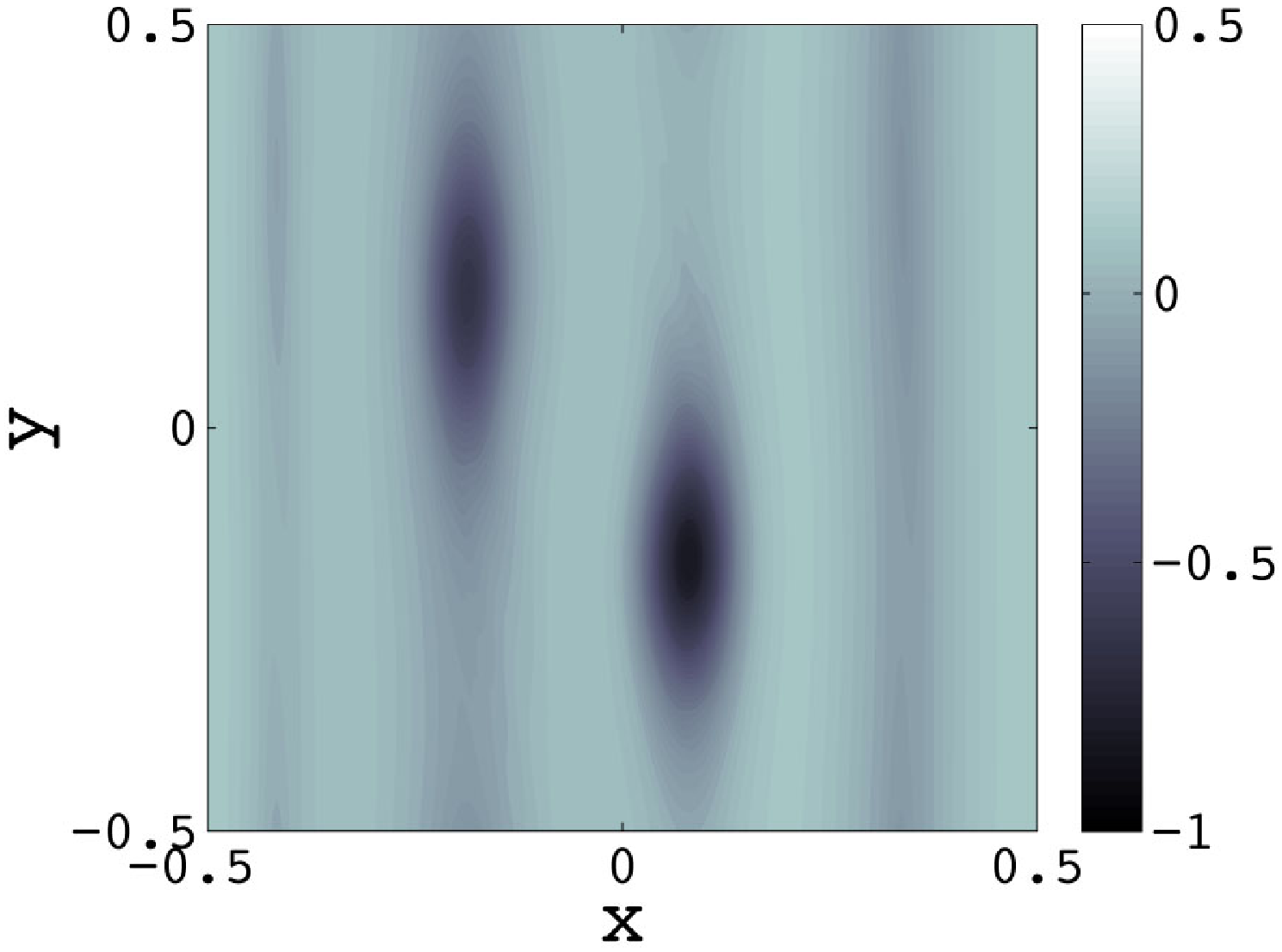}

   \caption{Evolution of the vorticity in the fiducial case. Top row has a baroclinic term with $Ri=-0.01$. Bottom row has no baroclinicity. We show $t=10$ (left), $t=100$ (middle), $t=500$ (right). }
              \label{snaps}%
\end{figure*}

When the system undergoes a subcritical transition, it develops long-lived \emph{self-sustained} vortices, as shown for $Ap=1$ in Fig.~\ref{snaps} (top). For such a large Reynolds number, it is known that vortices survive for long times \citep[see e.g.][]{UR04}, even without any baroclinic effect. To check that the observed vortices were really due to the baroclinic term, we have carried out the exact same simulation but without stratification (Fig.~\ref{snaps} bottom). This simulation also shows the formation of vortices but these are lately dissipated on a few hundred shear times. At $t=500\,S^{-1}$ the difference between the simulation with and without baroclinicity becomes obvious.

The vortices observed in these incompressible simulations lead to a very weak and strongly oscillating turbulent transport of angular momentum. One finds typically an \emph{inward} transport with $\alpha=\langle v_x v_y\rangle/SL^2\simeq -3\times 10^{-5}$. This is consistent with the results published by \cite{PSJ07} in global simulations.

In several other numerical experiments (not shown here), we have noticed that the amplitude threshold required to get the instability depends on $Re$ and $Pe$, a larger Reynolds number being associated with a weaker initial perturbation. This dependency was pointed out by \cite{PJS07} and it indicates that the amplitude threshold in a realistic disc could be very small (i.e. much smaller than the sound speed). Note however that this threshold might also be scale dependent, a problem which has not been addressed here.

\subsection{Influence of the Richardson number}
\begin{figure*}
   \centering
   \includegraphics[width=0.4\linewidth]{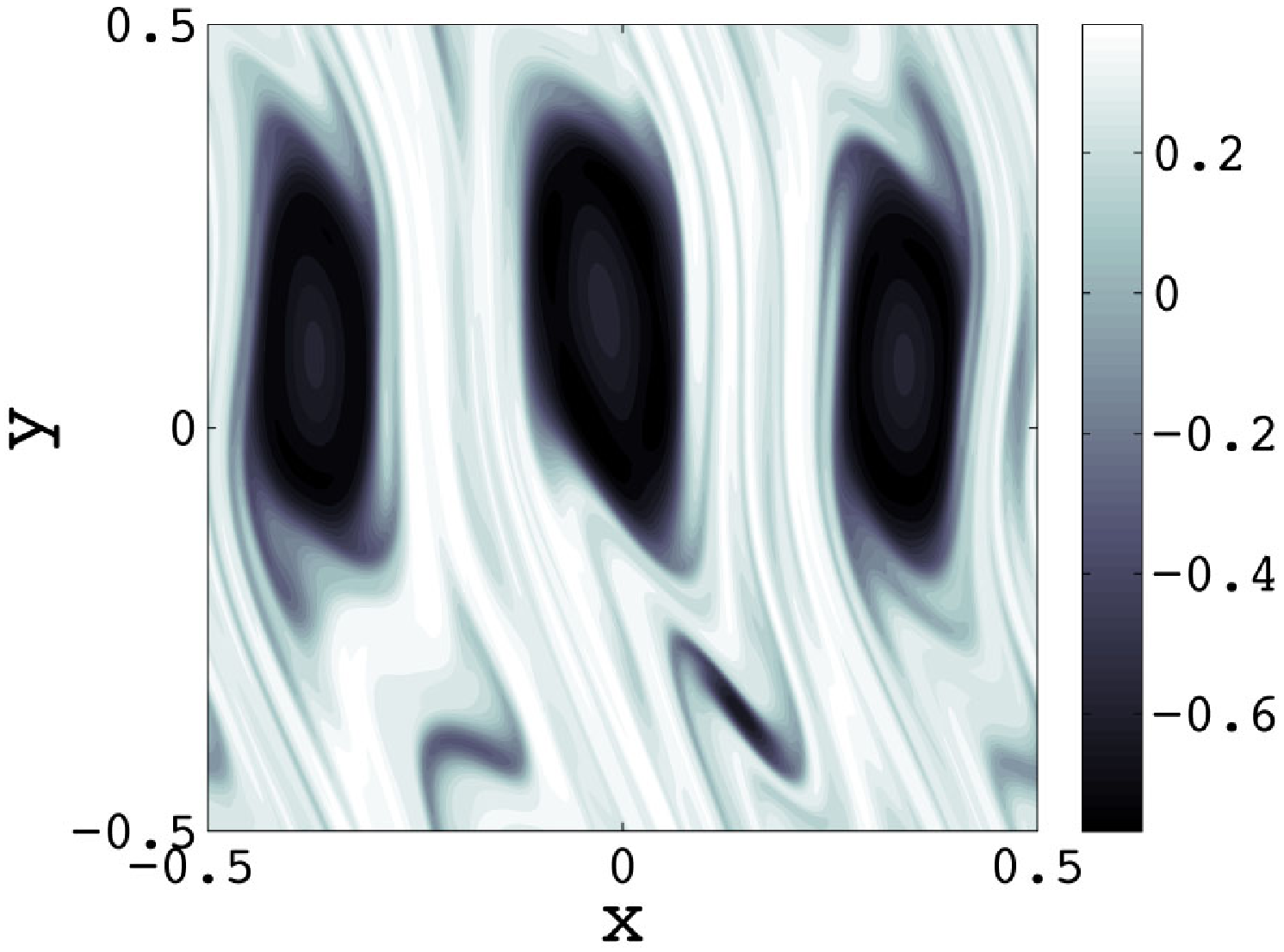}
   \quad\quad\quad
   \includegraphics[width=0.4\linewidth]{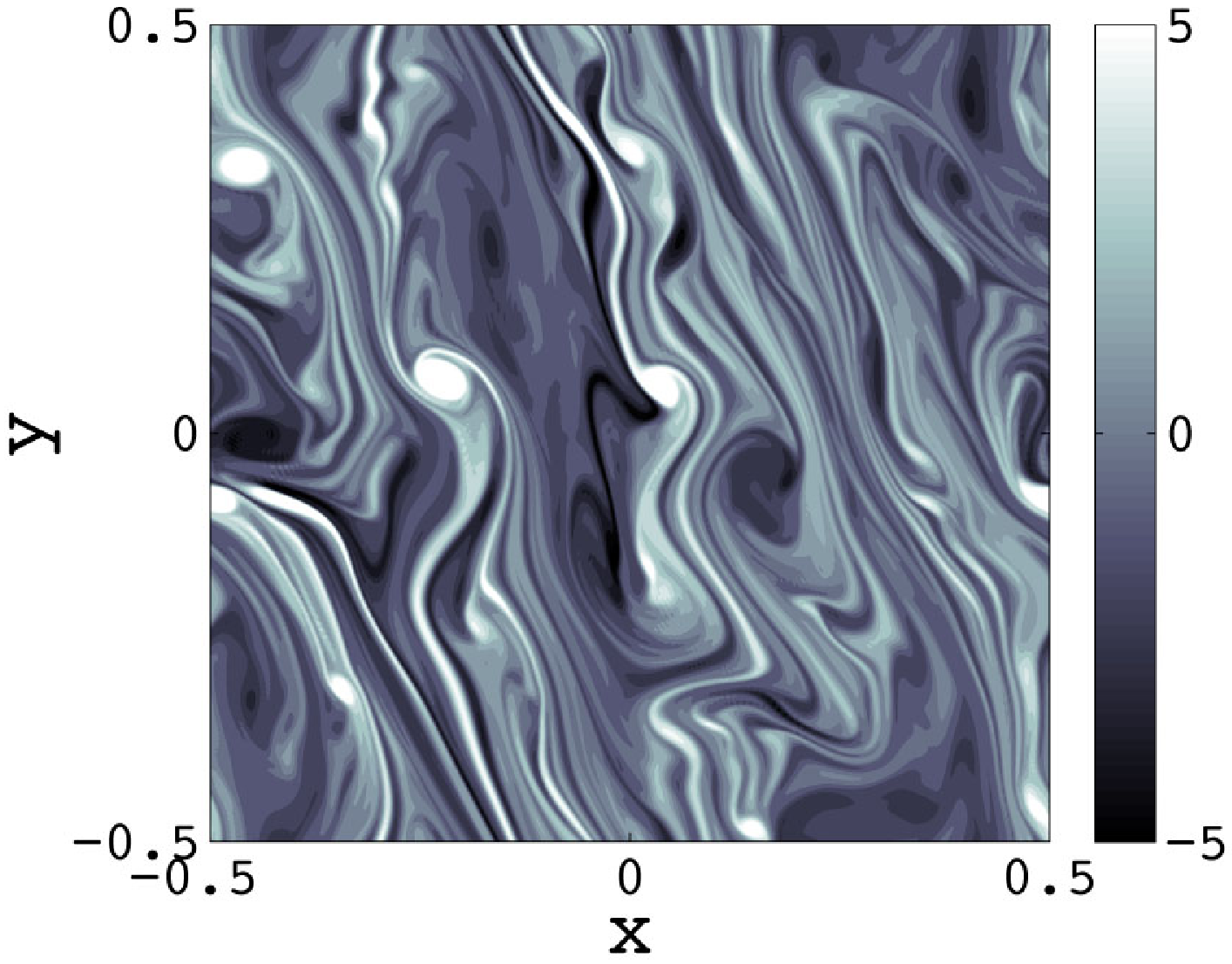}
   
   \caption{Vorticity map for $Ri=-0.02$ (left) and $Ri=-0.16$ (right) at $t=500$. As already shown by the spectra, small scales are dominant for larger $|Ri|$. }
              \label{Risnaps}%
\end{figure*}

\begin{figure}
   \centering
   \includegraphics[width=1.0\linewidth]{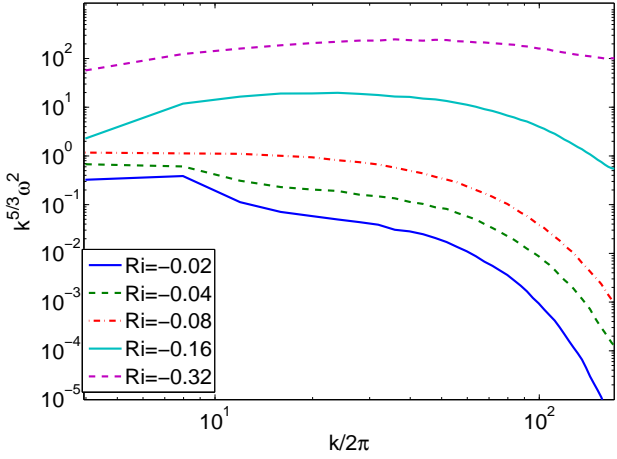}
   \caption{Enstrophy spectrum as a function of the Richardson number $Ri$. As $|Ri|$ becomes larger, the instability moves to smaller scales.}
              \label{spectrumN}%
\end{figure}

\begin{figure}
   \centering
   \includegraphics[width=1.0\linewidth]{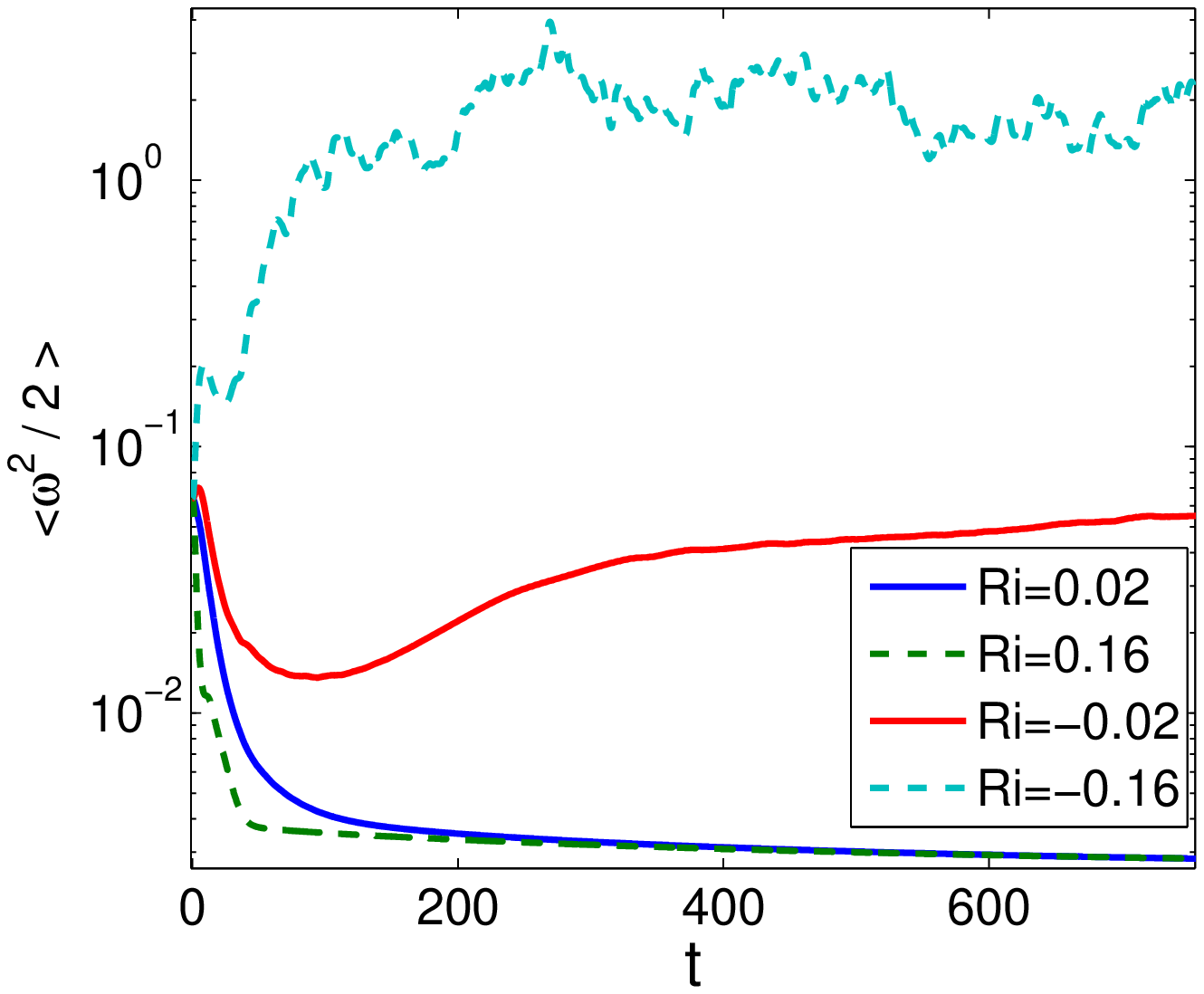}
   \caption{Time history of the total enstrophy for positive and negative Richardson number. The instability is observed only for negative $Ri$.}
              \label{posRi}%
\end{figure}

To understand how the instability depends on the amplitude of the baroclinic term, we have carried out several runs with $Re=4\times10^5$ and $Pe=4\times10^3$, varying $Ri$ from $-0.02$ to $-0.16$ and from $0.02$ to $0.016$. We compare the resulting shell-integrated enstrophy spectra ($\hat{\omega}^2_k/2$) on Fig.~\ref{spectrumN}. When $Ri<0$ and $|Ri|$ becomes larger, the instability tends to amplify smaller and smaller scales. In particular for $Ri=-0.02$, the dominant scale is clearly the box scale, as already observed for the fiducial run. This trend is also observed looking directly at vorticity snapshots (Fig.~\ref{Risnaps}). The sign of $Ri$ is also of importance for the onset of the SBI. To demonstrate this effect, we show the time history of the total enstrophy for positive and negative $Ri$ on figure \ref{posRi}. In the cases of positive $Ri$, the total enstrophy decays until the flow becomes axisymmetric. Perturbations are then damped very slowly on a viscous timescale. We conclude from these results that a necessary condition for the SBI to appear is a flow which do not satisfy the Schwarzschild criterion (\ref{schwarzschild}).

\subsection{Influence of thermal diffusion}
\begin{figure}
   \centering
   \includegraphics[width=1.0\linewidth]{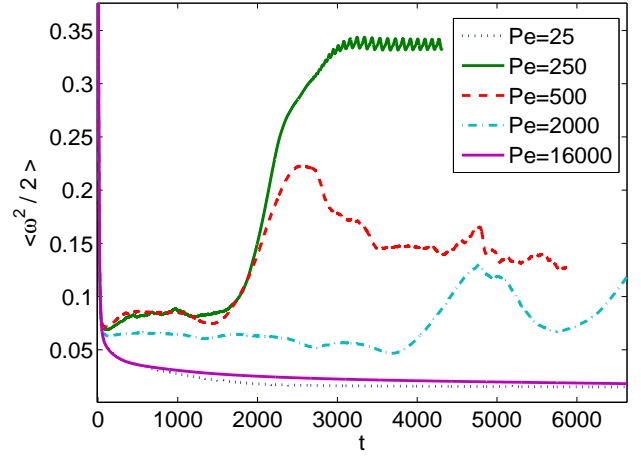}
   \caption{Time history of the total enstrophy for several Peclet number ($Pe$). The instability is observed for $50\le Pe \le 8000$.}
              \label{posPe}%
\end{figure}

The importance of thermal cooling and thermal diffusion was already pointed out by \cite{PSJ07}. To check this dependancy in our local model, we have considered several simulations with $Ri=-0.01$, varying $Pe$ from $Pe=20$ to $Pe=16000$. The resulting enstrophy evolution is presented for several of these runs on Fig.~\ref{posPe}. Looking at the snapshots of the vorticity field for these simulations, we find the SBI approximatively for $50\le Pe \le 8000$. Assuming a typical vortex size $l\sim 0.25$ (see e.g. Fig.\ref{snaps}), these Peclet numbers correspond to thermal diffusion times $3\,S^{-1}\le\tau_\mathrm{diff}\le 500\,S^{-1}$ over the vortex size, with an optimum found for $\tau_\mathrm{diff}\simeq 10\,S^{-1}$ ($Pe=250$). Note that the cooling time used by \cite{PSJ07} lies typically in this range of values.

\subsection{Instability mechanism}
Since the flow is subcritically unstable, no linear analysis can capture this instability entirely. As shown by \cite{JG05}, an ensemble of shearing waves in a baroclinic flow is subject to a transient growth  with an amplification going asymptotically like\footnote{Note that the amplification occurs independently of the sign of $Ri$, as already mentioned by \cite{JG05}.} $|Ri|t^{1-4Ri}$ for $|Ri|\ll 1$ when $\mu=\nu=0$, the waves being ultimately decaying when $\mu>0$ or $\nu>0$. However, this tells us little to nothing about the nonlinear behaviour of such a flow. Indeed, it is known that barotropic Keplerian flows undergo arbitrarily large transient amplifications \citep[see e.g.][]{CZTL03}, but subcritical transitions are yet to be found in that case \citep{HBW99,LL05,JBSG06}. In the SBI case, the subcritical transition happens only for negative $Ri$ as shown above, in flagrant contradiction with the linear amplification described by \cite{JG05}. This is a strong indication that linear theory is of little use for describing this instability.

\begin{figure*}
   \centering
   \includegraphics[width=0.4\linewidth]{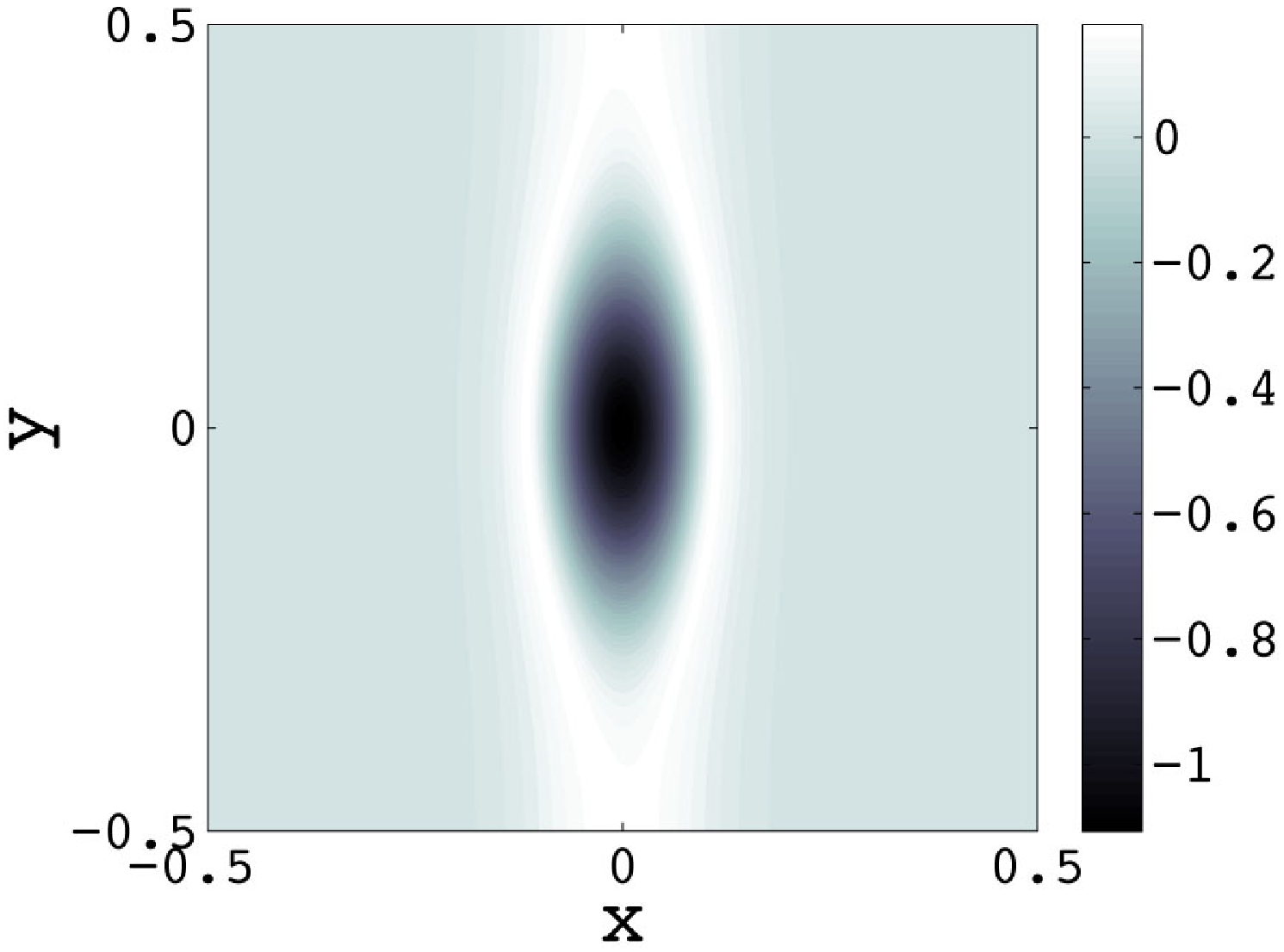}
   \quad\quad\quad
   \includegraphics[width=0.4\linewidth]{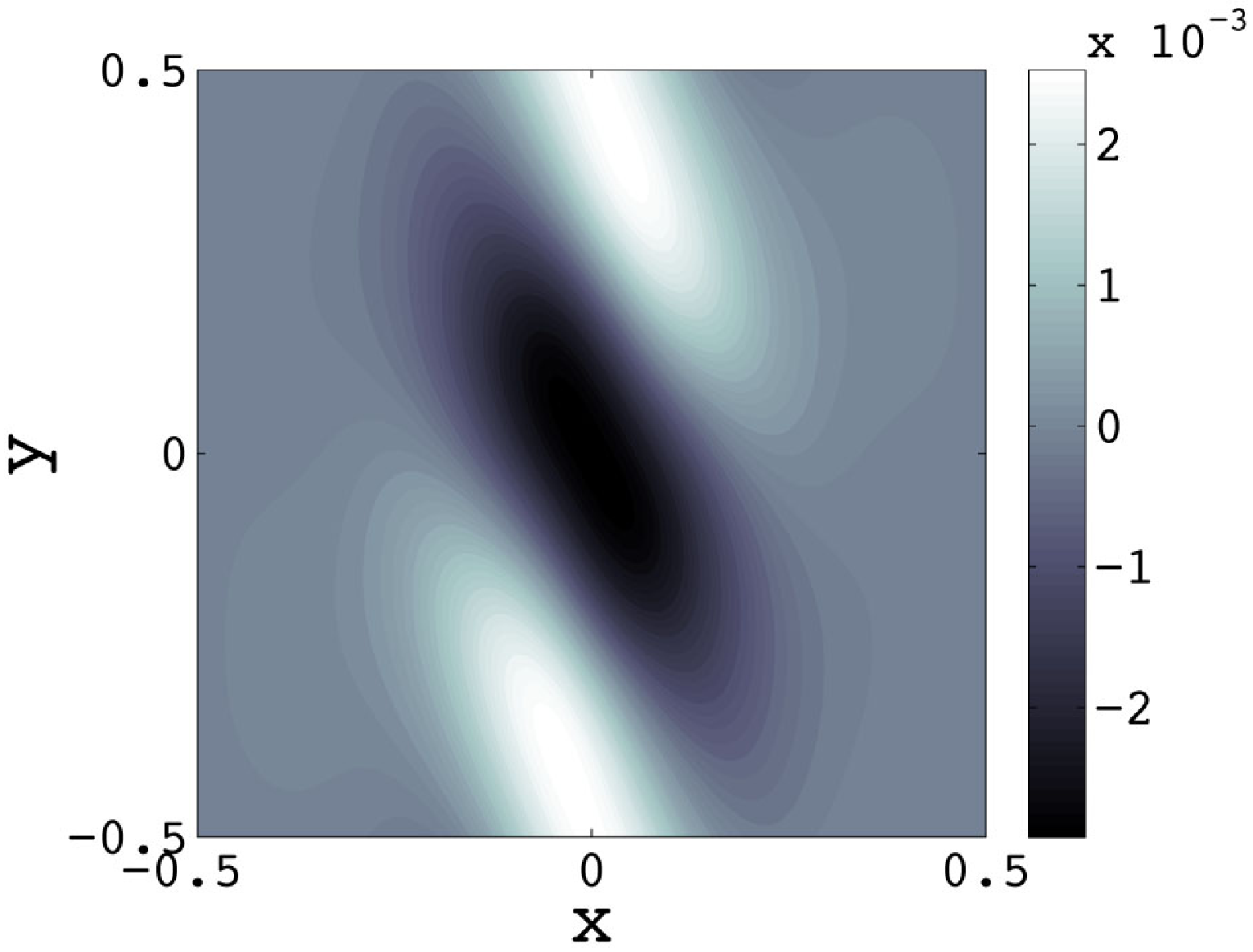}
   
   \caption{Vortex structure obtained starting from a Kida vortex. (left) vorticity map. (right) baroclinic term ($\Lambda N^2\partial_y \theta$) map.}
              \label{Kida-struct}%
\end{figure*}

To isolate a mechanism for a subcritical instability, one should start from a non-linear structure observed in the simulations. For the SBI, these structures are self-sustained vortices. In order to understand how baroclinic effects can feedback on the vortex structure, we initialised our fiducial simulation with a Kida vortex of aspect ratio 4 \citep{K81}. Although this vortex is known to be an exact nonlinear solution of the inviscid equation of motions, it is slowly modified by the explicit viscosity and the baroclinic term, leading to a slow growth of the vortex. We show on Fig.~\ref{Kida-struct} the resulting vortex structure (left) and the associated baroclinic term $\Lambda N^2\partial_y \theta$ (right) at $t=100\,S^{-1}$. Note that these structures are quasi steady and evolve on timescales much longer than the shear timescale. It is clear from these snapshots that the baroclinic feedback tends to amplify the vorticity located inside the vortex, leading to the growth of the vortex itself.

To understand the origin of the baroclinic feedback, let's assume the physical quantities are evolving slowly in time, 
as is observed in the simulations, so we may write
\begin{eqnarray}
\omega &=& \omega(\epsilon t)\equiv\omega(\tau),\\
\theta &=& \theta(\epsilon t)\equiv \theta(\tau),
\end{eqnarray}
where $\epsilon$ is a small parameter and it is assumed that $|\theta_{\tau}|$ is of the same order as $|{\bf v}\cdot\nabla \theta |$, and similarly for $\omega$.  We therefore have to solve:
\begin{eqnarray}
\epsilon\partial_{\tau}\omega+(\bm{v\cdot\nabla}-Sx\partial_y)\omega&=&\Lambda N^2\partial_y\theta+\nu\Delta\omega,\\
\label{eq_vort_0}\epsilon\partial_{\tau}\theta+(\bm{v\cdot\nabla}-Sx\partial_y)\theta&=&v_{x}/\Lambda +\mu\Delta\theta.
\end{eqnarray}
Since the Richardson number is assumed to be small, the baroclinic feedback of (\ref{eq_vort_0}) has to be small. As only this term can lead to an instability, we can assume it scales like $\epsilon$. The role of the viscosity is to prevent the instability from happening, damping vorticity fluctuations. Assuming we are in a regime in which the instability appears, the viscosity has to be of the order of the baroclinic term, scaling like $\epsilon$. At zeroth order in $\epsilon$, we are left with:
\begin{eqnarray}
\label{invisid_0}(\bm{v\cdot\nabla}-Sx\partial_y)\omega&=&0,\\
\label{entrop_0}(\bm{v\cdot\nabla}-Sx\partial_y)\theta&=&v_{x}/\Lambda+\mu\Delta\theta
\end{eqnarray}
This system of equations describes a quite simple physical system: the vorticity field is a steady solution of the inviscid vorticity equation with a constant shear and without any baroclinic feedback, whereas the potential temperature results from advection-diffusion of the background entropy profile due to the flow structure. A family of steady solutions of the inviscid vorticity equation with a constant shear are known to be steady vortices, like the \cite{K81} vortex. More generally, it can be shown that any $\omega$ of the form:
\begin{equation}
\omega=F(\psi)
\end{equation}
where $\psi$ is the stream function defined by $\bm{u}=\bm{e_z \times \nabla} \psi$ is solution of (\ref{invisid_0}). 

In the following, we assume (\ref{invisid_0}) is satisfied by $\omega$ and we look for a solution to the entropy equation (\ref{entrop_0}). As a further simplification, we neglect the advection of the perturbed entropy $\theta$ by $\bm{v}$, assuming this effect is small compared to the advection by the mean shear, although this  does not affect the argument
leading to the possibility of instability (see below). Using a Fourier decomposition $\theta(\bm{x})=\int d\bm{k}\,\hat{\theta}(\bm{k})\exp(i\bm{k\cdot x})$, one gets
\begin{equation}
\label{adv_diff}Sk_y\frac{d\hat{\theta}}{dk_x}+\mu k^2\hat{\theta}=\hat{v}_x/\Lambda .
\end{equation}
A solution to this equation can easily be found using standard  techniques for solving first order ordinary differential equations. Assuming $|\hat{v}_x|$ decays to zero when $|\bm{k}|\rightarrow\infty$, one may  write the result in the form
\begin{equation}
\label{ent_struct}\hat{\theta}(k_x,k_y)=\frac{1}{\Lambda |Sk_y|}\int_{-\infty}^\infty dk'_x \hat{v}_x(k'_x,k_y)G(k_x,k_x',k_y),
\end{equation}
where we have defined
\begin{eqnarray}
\nonumber G(k_x,k_x',k_y)&=&\mathcal{H}\Big((k_x-k'_x)Sk_y/|Sk_y|\Big) \times\\
& & \exp\Big[\frac{\mu}{Sk_y}\Big(\frac{k_x'^3-k_x^3}{3}+k_y^2[k_x'-k_x]\Big)\Big],
\end{eqnarray}
$\mathcal{H}$ being the Heaviside function. From this expression, it is formally possible to derive the time evolution of the vorticity and look for an instability using the first order term in (\ref{eq_vort_0}). However, this requires an explicit expression for the vorticity, which is \emph{a priori} unknown. Instead, we have choosen to compute the evolution of the total enstrophy of the system $\langle \omega^2/2\rangle=\int dxdy \, \omega^2/2$:
\begin{equation}
\label{final_vort_budget}
\partial_t\langle \omega^2/2 \rangle=\Lambda N^2\langle\omega\partial_y\theta\rangle-\nu\langle|\bm{\nabla} \omega|^2 \rangle
\end{equation}
Evidently, an instability can occur only if the first term on the right hand side is positive. Using (\ref{ent_struct}), is is possible to derive an analytical expression for this term: 
\ \

\ \

\ \

\ \
\begin{eqnarray}
\nonumber \Lambda N^2\langle\omega\partial_y\theta\rangle&=&  -4\pi^2 \Re\Bigg[N^2\int dk_x dk_y dk'_x\\
\label{BCfeedback}& &\frac{|k_y|}{|S||{\bm k'}|^2} \hat{\omega}^*(\bm{k})G(k_x,k'_x,k_y)\hat{\omega}(\bm{k'})\Bigg],
\end{eqnarray}
where ${\bm k'}= (k_x',k_y).$
In this expression, the existence of an instability is controlled by the Green's function $G(k_x,k'_x,k_y)$ and the shape of
 $\omega(\bm{k})$. To our knowledge, it is not possible to derive any general criterion for the instability, as such a criterion would depend on the background vortex considered. It is however possible to derive a criterion in the limit of a large thermal diffusivity $\mu$. In this limit, $G$ is strongly peaked around $k'_x=k_x$, and one can assume in first approximation that $G$ is proportional to a Dirac distribution. More formally, one can approximate in the limit of large thermal diffusivity:
\begin{equation}
G(k_x,k_x',k_y)\simeq\mathcal{H}\Big((k_x-k'_x)Sk_y/|Sk_y|\Big)\exp\Big[\frac{\mu}{Sk_y}k^2(k_x'-k_x)\Big],
\end{equation}
when $k_x\simeq k_x'$. It is then possible to derive an expression for the baroclinic feedback as a series expansion in the small parameter $\mu^{-1}$:
\begin{eqnarray}
\nonumber LN^2\langle\omega\partial_y\theta\rangle&\simeq& -4\pi^2\Re\Big[N^2\int d\bm{k}\Big(\frac{k_y^2}{\mu k^4}|\hat{\omega}(\bm{k})|^2-\\
& &\frac{Sk_y^3}{\mu^2 k^4}\hat{\omega}^*(\bm{k})\partial_{k_x}\big[\hat{\omega}(\bm{k})/k^2\big]+\dots \Big)\Big]\label{growth}
\end{eqnarray}
In this approximation, we find two competing effects. The first term on the right hand side appears when the thermal diffusion completely dominates over the shear in (\ref{adv_diff}). It has a positive feedback on the total enstrophy when $N^2<0$ and can be seen as a source term for the SBI. The second term involves a competition between shear and diffusion. It can be seen in first approximation as a phase mixing term, and the resulting can be either positive or negative, depending on the vorticity background one considers. Physically, this term shears out the entropy structure created by the vortex and if it is too strong, it kills the positive feedback of the first term. One should note however that this term will not \emph{reverse} the sign of the baroclinic feedback but will just weaken it by randomising the phase coherence between $\omega$ and $\theta$.

To understand the ``growth'' described by (\ref{growth}), on can derive an order of magnitude expression for the growth rate $\gamma=d\ln\,\langle \omega^2\rangle/dt$ combining (\ref{final_vort_budget}) with (\ref{growth}):
\begin{equation}
\gamma\sim\frac{(-N^2)\sigma^2}{\mu}\phi_\omega (S\sigma^2/\mu)-\frac{\nu}{\sigma^2}.
\end{equation}
In this expression, $\sigma$ is the typical vortex size, with the assumption $\sigma\sim 1/k$. We have included a phase mixing term with the function $\phi_\omega$, as an extension of (\ref{growth}). This function depends on the background vortex solution $\omega(x,y)$ one chooses and cannot be written explicitly in general. As explained above, this phase mixing weaken the baroclinic feedback when $Pe$ is large but it does not change its intrinsic nature. According to these properties, one expect $\phi_\omega(0)=1$ and $\phi_\omega(x) \rightarrow 0$ when $x\rightarrow \infty$.

Although $\phi_\omega$ can only be determined for a specific vortex solution, one can still deduce a few general properties for the SBI. For very large thermal diffusivity (very small $Pe$) one will expect the SBI when 
\begin{equation}
\frac{(-N^2)\sigma^4}{\nu\mu} > 1,
\end{equation}
since $\phi_\omega\sim 1$ in this limit. Although not quantitatively equivalent, this first limit corresponds to the $Pe>25$ threshold of our simulations. It is moreover similar to the criterion for the onset of convection (in the absence of shear) based on the Rayleigh number $Ra=-N^2\sigma^4/\nu\mu$. One finds another instability condition due to phase mixing when $\mu\rightarrow 0$. Assuming $\phi_\omega(x)$ decays faster than $1/x$ when $x\rightarrow\infty$, one gets a minimum value for $\mu$ solving
\begin{equation}
\frac{\phi_\omega(S\sigma^2/\mu)}{\mu}>\frac{\nu}{(-N^2) \sigma^4}
\end{equation}
which can be calculated once $\phi_\omega$ is explicitly provided. In our simulations, this second limit corresponds to the $Pe<16000$ threshold. 

Although not entirely conclusive, this short analytical analysis tends to explain why a relatively strong thermal diffusion is required to get the SBI. This dependancy was also pointed out by \cite{PSJ07} who used both a thermal diffusion and a cooling relaxation time in the entropy equation. Such a cooling time can also be introduced in our analysis in place of the thermal diffusion but this does not change the underlying physical process. We also note that in this analysis, we have assumed that a finite amplitude background vorticity satisfying (\ref{invisid_0}) existed in the first place, from which we have derived the baroclinic feedback on the same vorticity field. In this sense, this analysis describes a non-linear instability and is different from the linear approach of \cite{JG05}.  In addition, we recall that we made 
the approximation of only including the background shear in the advection term 
on the left hand side of  equation (\ref{entrop_0}).   We remark that
 the dominant driving term in the limit
of large $\mu$ in (\ref{growth}) does not depend on this assumption because the heat diffusion term dominates
in this limit.

\subsection{Phenomenological picture}
\begin{figure}
   \centering
   \includegraphics[width=1.0\linewidth]{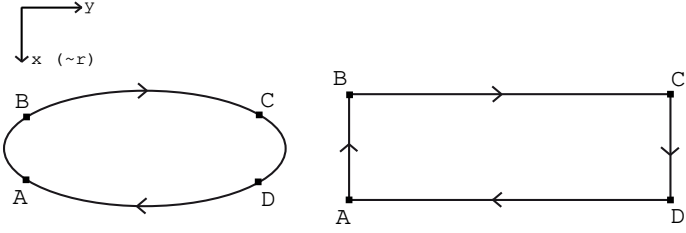}
   \caption{Streamline in a vortex undergoing the SBI. A fluid particle is accelerated by buoyancy effects on A-B and C-D branches. Cooling occurs on B-C and D-A branches (see text).}
              \label{streamline}%
\end{figure}

Knowing the basic mechanisms underlying the SBI, one can tentatively draw a phenomenology of this instability. For this exercise, let us consider a single streamline in a vortex subject to the SBI (Fig.~\ref{streamline} left). For simplicity, we reduce the trajectory followed by a fluid particle in the vortex to a rectangle (Fig.~\ref{streamline} right). Let us start with a fluid particle on A, moving radially inward toward B. This fluid particle is initially in thermal equilibrium, and we assume the motion from A to B is fast enough to be considered approximately adiabatic. As the particle moves inward, its temperature and density slowly deviate from the background values. If the background entropy profile is convectively unstable (condition \ref{schwarzschild}), the fluid particle moving inward is cooler and heavier than the surrounding, and is consequently subject to an inward acceleration due to gravity (the particle ``falls''). Once the particle reaches B, it drifts azimuthally toward C. On this trajectory, the background density and temperature is constant. Therefore, the fluid particle gets thermalised with the background and, if the cooling is fast enough, it reaches C in thermal equilibrium. Between C and D, the same buoyancy effect as the one between A and B is observed: as the particle moves from C to D, it is always hotter and lighter than the surrounding, creating a buoyancy force directed radially outward on the particle. Finally, a thermalisation episode happens again between D and A, closing the loop.

From this picture, it is evident that fluid particles get accelerated by buoyancy forces on A-B and C-D branches. In the end, considering many particles undergoing this buoyancy cycle, the vortex structure itself gets amplified, explaining our results. Moreover, the role played by cooling or thermal diffusion is evident. It the cooling is too fast, fluid particles tend to get thermalised on the A-B and C-D branches, reducing the buoyancy forces and the efficiency of the cycle. On the other hand, if no cooling is introduced, the particles cannot get thermalised on the B-C and D-A branches. In this case, the work due to buoyancy forces on the B-A trajectory will be exactly opposite to the work done on the C-D trajectory, neutralising the effect on average.

\section{2D SBI in compressible flows\label{compressibility}}

In this situation we consider the 2D subcritical baroclinic instability 
within the framework of  a compressible model.
We accordingy adopt the compressible counterparts of  equations (\ref{motiongeneral})-
(\ref{divv}) for an ideal gas  with constant
ratio of specific heats $\gamma$ in the form
\begin{eqnarray}
\nonumber  \frac{D \bm{u}}{Dt}&=&-\frac{1}{\rho}\bm{\nabla}P -\bm{\nabla}\Phi
-2\bm{\Omega \times u}\\
\label{motiongeneralcomp}& &-2\Omega S x \bm{e_x}+\nabla\cdot\bm{T_v},\\
\label{entropygeneralcomp}\frac{D c^2}{Dt}  &=&
 \frac{(\gamma-1)c^2}{\rho}\frac{D \rho}{Dt} +
\frac{1}{\rho}\left({\cal H} + {\cal K}\Delta c^2\right)\\
\label{divvcomp}\partial_t\rho  &=&-\bm{\nabla \cdot\rho u}.
\end{eqnarray}
Here $c$ is the isothermal sound speed which
is proportional to the square root of the  temperature,
 $\bm{T_v}$ is a viscous stress tensor,
${\cal H}/(\gamma - 1)$ is the rate of energy production per unit volume, $\Phi$ is a general external
potential, and the constant  $K = \mu \rho$ where as for
the incompressible case $\mu$ is a thermal diffusivity.
\begin{figure}
   \centering
   \includegraphics[width=0.9\linewidth]{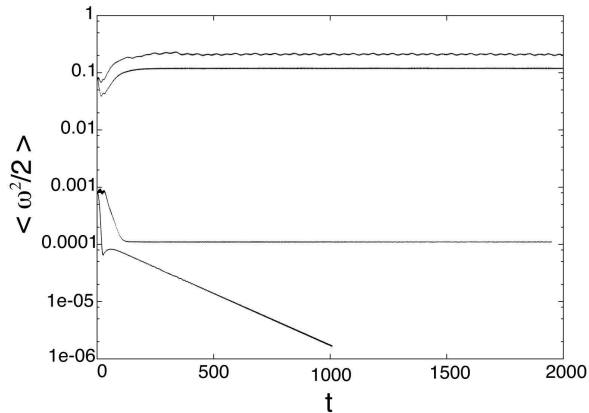}
   \caption{Time history  of the evolution of the enstrophy  corresponding to the
   perturbations for the small box. The uppermost
   curve is for ${\cal A}=0.25$   with no applied viscosity. The next uppermost
   is the corresponding  case  with an imposed viscosity. The lowermost curve is for
   ${\cal A}=0.025$ with applied viscosity and the next lowermost curve is for the corresponding
   case with no viscosity}
              \label{smallboxenst}%
\end{figure}

\begin{figure}
   \centering
   \includegraphics[width=0.9\linewidth]{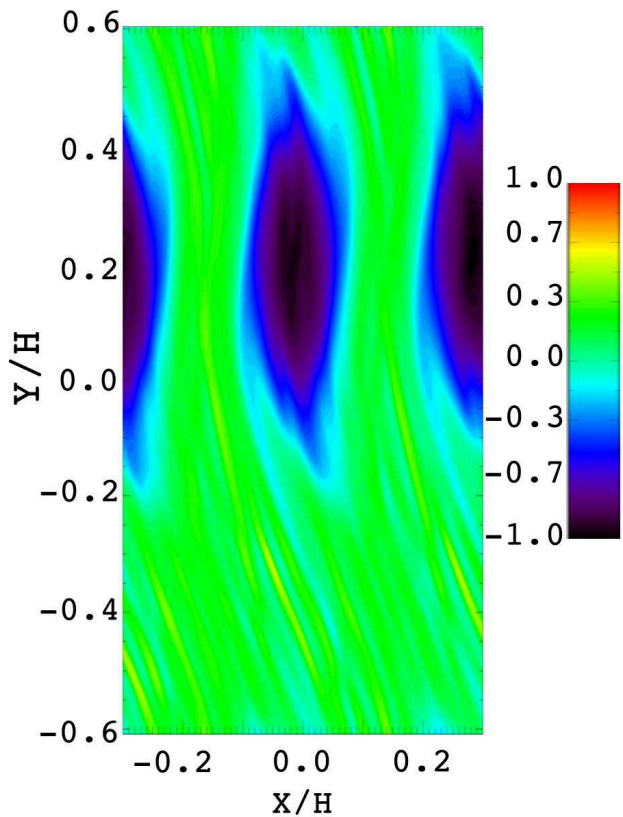}
   \caption{Vorticity map for the small box simulation with applied viscosity
   and ${\cal A}=0.25$  at $t=2000.$}
              \label{smallboxvisc}%
\end{figure}

For simplicity we adopt a  model that may be either regarded as assuming
that there is a prescribed energy production rate ${\cal H}$
and $\gamma$ is close to unity in (\ref{entropygeneralcomp}) so that the compressional
heating term $\propto (\gamma-1)$ can be dropped, or  replacing the combination of the compressional heating term and ${\cal H}$ by a new specified energy production / cooling rate.
In either view, as the energy production rate is specified, the model is simplified as
 equation (\ref{entropygeneralcomp}) becomes an equation for $c^2$ alone.
 
We consider shearing box models  of extent $L_x$ in the radial direction and $L_y$ in the azimuthal
direction. A characteristic constant  sound speed, $c_0,$  defines a natural scale height $H=c_0/\Omega.$
We present simulations  using NIRVANA (see section \ref{Numerics}) 
for a small box with $L_x=L_y/2=0.6H$ and a large box with  $L_x=L_y/2=1.2H.$
Thus in terms of the characteristic sound speed,  when Keplerian shear
is the only motion,  relative motions in the small box are subsonic, whereas supersonic relative velocities
occur in the big box. We have also considered  the two resolutions $(N_x,N_y)=(144,288)$ and $(N_x,N_y)=(288,576)$
and tests have shown that  these give the same results.  We have  also performed simulations with no applied viscosity
and with an applied viscosity corresponding to a Reynolds number $L_x^2\Omega/\nu =  12500.$
This was applied as a constant kinematic  Navier Stokes viscosity but with stress tensor acting on the velocity
${\bf v}$ being the deviation from the background shear rather than ${\bf u}.$
  This is not significant for an incompressible simulation but ensures that unwanted phenomena such as viscous overstabilty
  produced by perturbations of  the background viscous stress \citep[see][]{KPL93}
  are absent. The diffusivity corresponding
  to the viscosity we used is the same magnitude as the magnetic diffusivity used in \cite{FPLH07} 	
  and as in their case  we find that it is adequately resolved. 

\begin{figure}
   \centering
   \includegraphics[width=0.9\linewidth]{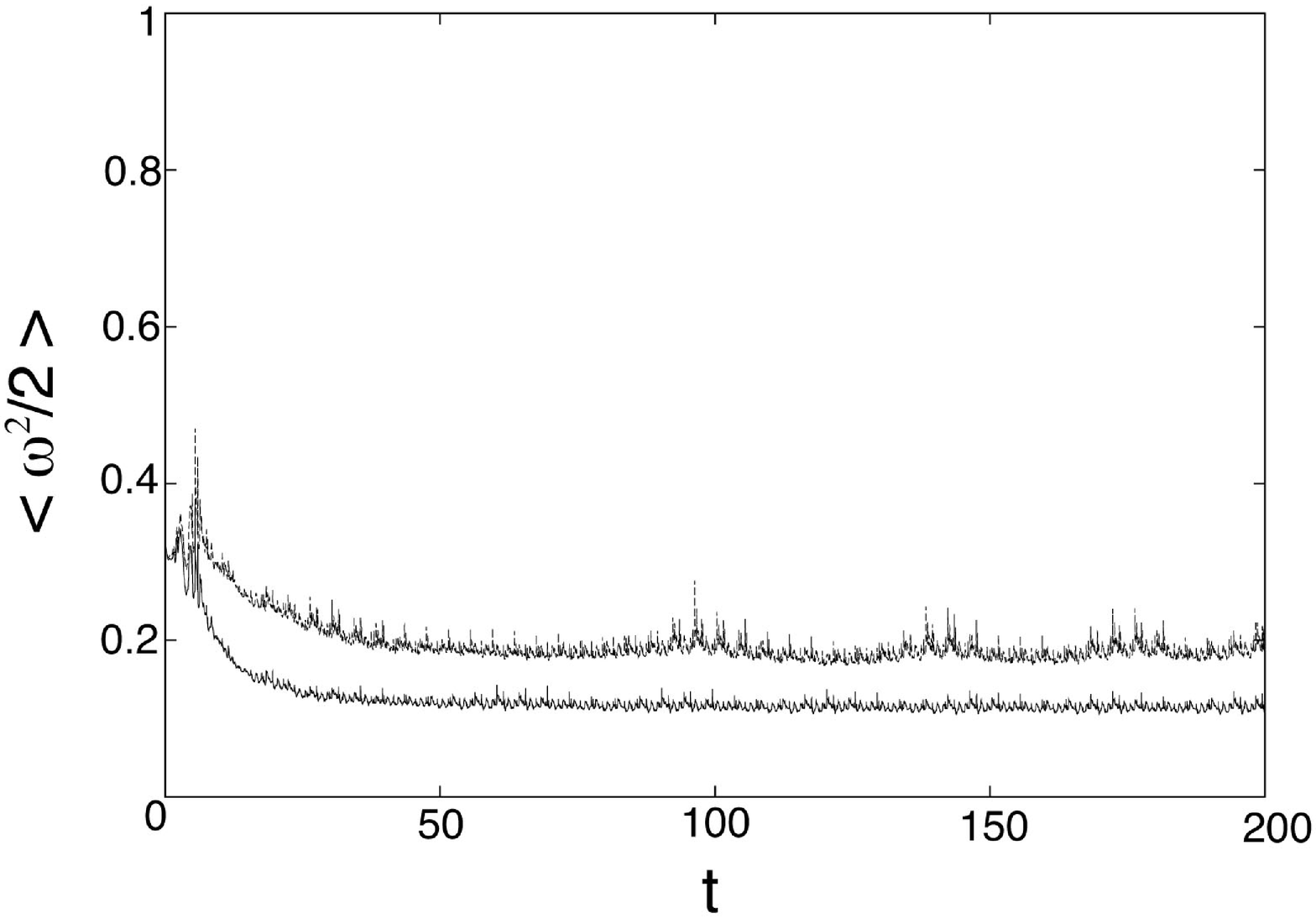}
   \caption{Time history of the evolution of the enstrophy corresponding
   to the perturbations for the large box. The uppermost
   curve is for ${\cal A}=0.5$   with no applied viscosity. The lower curve
   is for the corresponding  case  with an imposed viscosity. }
              \label{Bigboxenstrophy}%
\end{figure}

\begin{figure*}
   \centering
   \includegraphics[width=0.4\linewidth]{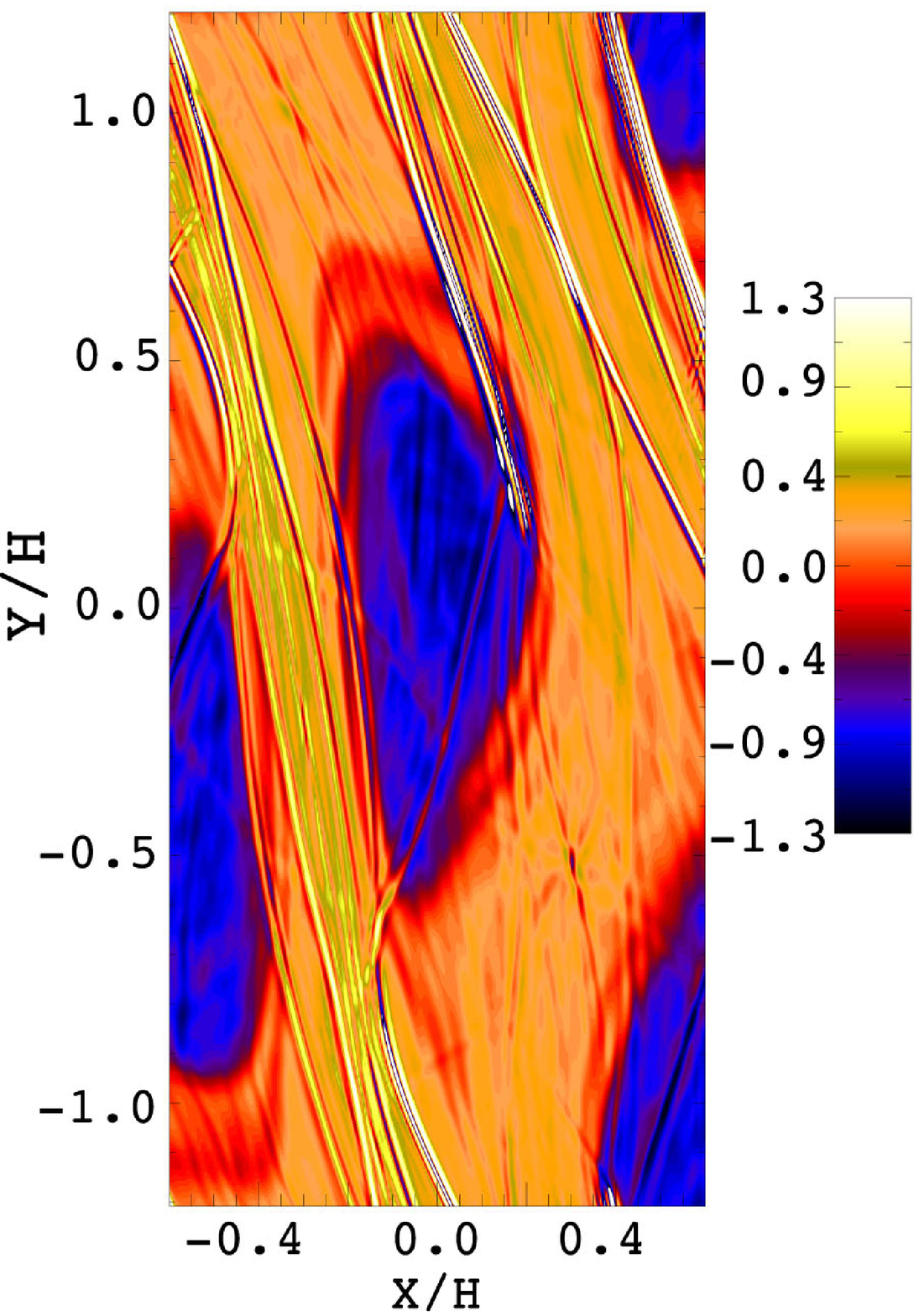}
   \quad\quad\quad\quad
   \includegraphics[width=0.4\linewidth]{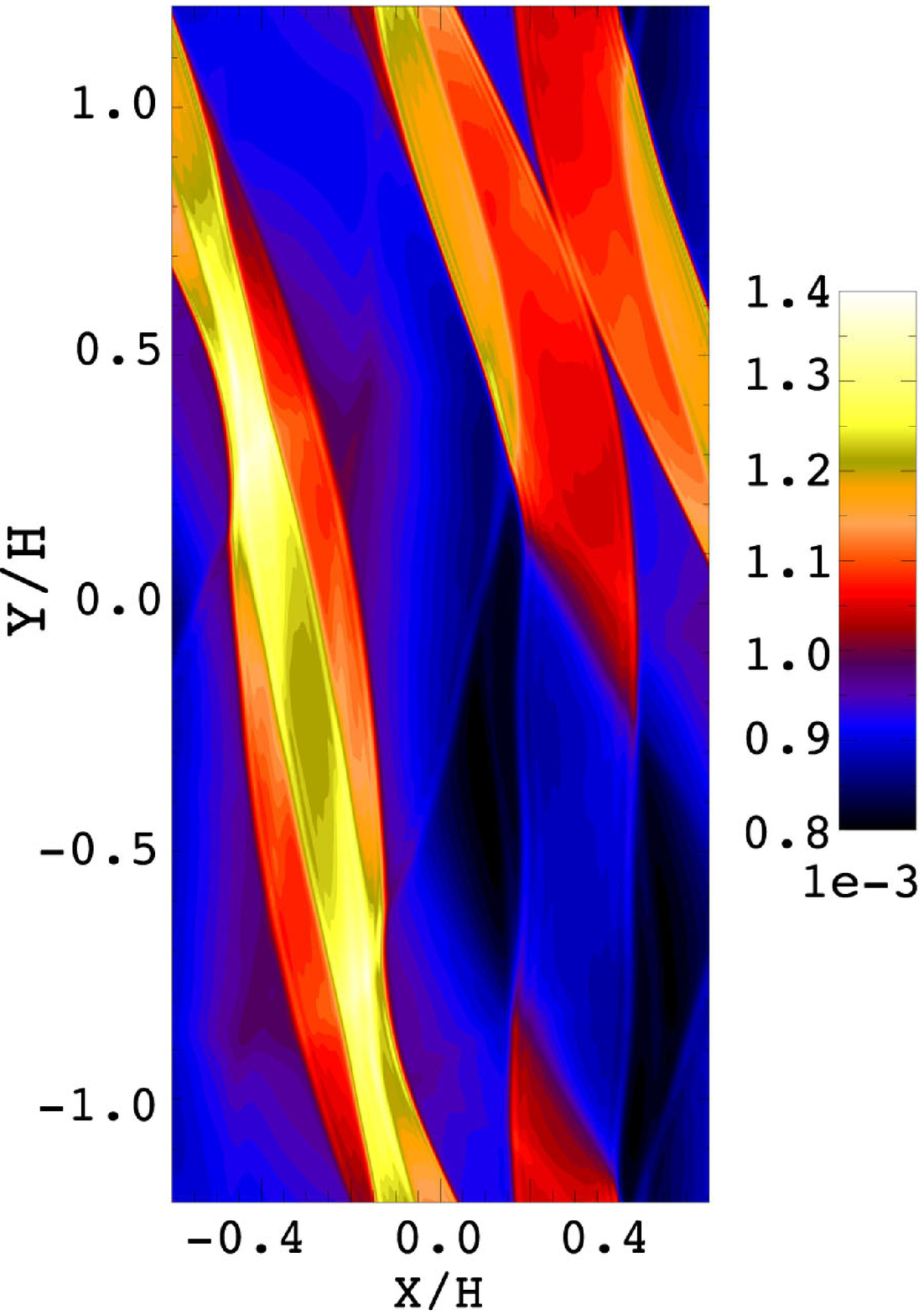}
   \caption{Vorticity map (left) and surface density map (right) for the large box simulation with applied viscosity
   and ${\cal A}=0.5$ at t=190.}
              \label{bigboxviscfig}%
\end{figure*}
      
In order to perform simulations corresponding to the incompressible ones 
we need to impose gradients in the state variables that produce a non zero Richardson
number (appropriate for $\gamma =1.$)
\begin{equation}
Ri =-\frac{1}{ S^2 \rho}\frac{\partial P}{\partial R}\frac{\partial}{\partial R}\ln\Big(\frac{P}{\rho}\Big).
\end{equation}
Because of the shearing box boundary conditions,  imposed background gradients
have to be periodic. Thus we choose  an initially uniform density $\rho= \rho_0$
and initial profiles for $P$ and $c^2$ of the form
$P =\rho_0c_0^2 (1- C_p\sin(2\pi x/L_x))$  and $c^2 =c_0^2(1 -C_p\sin(2\pi x/L_x)),$
where $C_p$ is a constant.
For a given value $H/L_x,$ $Ri$ depends only on $C_p$ which was chosen to
give a maximum value for this quantity of $-0.07.$ Note that in our case, as $C_p$ is small,
we have approximately $Ri\propto \cos^2(2\pi x/L_x)$
in contrast to the incompressible simulations where it is constant.  Once the initial value of  $\mu$
has been chosen to correspond to a specified  uniform Peclet number, the external potential $\Phi$
and the energy source term ${\cal H}$ were chosen such that the choice of state variables
with Keplerian shear resulted in an equilibrium state and then held fixed. For comparison with the incompressible
simulations we adopted $Pe=379$ for the simulations reported here.
At this point we stress that in view of the rather artificial set up, this model is clearly far from definitive.
However, it can be used to show that the phenomena found in the incompressible case
are also seen in a compressible model but with the addition of the effects of density waves
in the case of the large box.

As for the incompressible case, solutions with sustained vortices were 
only found for sufficiently large  initial velocity perturbations of the equilibrium state.
We adopted an incompressible velocity perturbation of the form
\begin{equation}
v_x = (3L_y\Omega /(8\pi)){\cal A}\sin(2\pi y/L_y)\cos(2\pi x/L_x)\end{equation}
and 
\begin{equation}
 v_y =- (3L^2_y\Omega /(8\pi L_x)){\cal A}\cos(2\pi y/L_y)\sin(2\pi x/L_x),
 \end{equation}
where ${\cal A}$ is specified amplitude factor.

\subsection{Simulation results}

\begin{figure}
   \centering
   \includegraphics[width=0.9\linewidth]{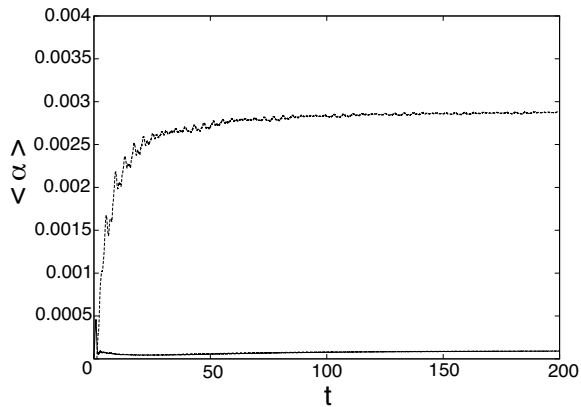}
   \caption{Time history of the running means of the 
   the spatially averaged value of  $\alpha$ 
   for the small box (lower curve) and large box (upper curve)
   simulations with applied viscosity and
   with  ${\cal A}=0.5$ and ${\cal A}=0.25$ respectively.}
              \label{alphaplot}%
          \end{figure}

   In Fig. \ref{smallboxenst} we show the
     the evolution of the enstrophy associated with the
   perturbations for the small box. Cases with
    ${\cal A}=0.25$   with and without  applied viscosity are shown.
   These lead to solutions for which anticyclonic vortices are sustained
   for long times. The case with no viscosity is more active as expected
   but otherwise looks similar to the case with applied viscosity.
   By contrast when the amplitude of the initial perturbation is reduced
   to ${\cal A}=0.025$ no sustained vortices are seen. 
   This demonstrates that our numerical setup is not subject to any
   linear instability, such as Rossby wave instabilities \citep{LLC99}.
   The enstrophy attains
   a low level 
   in the  case with no applied viscosity which is a consequence
   of a long wavelength linear axisymmetric disturbance that shows
   only very weak  decay provided by numerical viscosity in this run.
   Fig. \ref{smallboxvisc} shows a vorticity map for the small box simulation
   with applied viscosity and ${\cal A}=0.25$. Anticylconic vortices are clearly 
   seen in this case supporting the finding from the incompressible
   runs that a finite amplitude initial kick is required to generate them.

    The time history of the evolution of the enstrophy for 
    large box simulations with  ${\cal A}=0.5$   with and
    without  applied viscosity is shown in  Fig. \ref{Bigboxenstrophy}.
    Corresponding vorticity  and surface density maps for the
    case with applied viscosity are  shown in  Fig. \ref{bigboxviscfig}.
    Again the inviscid case is more active but nonetheless the corresponding
    maps look very similar. In these cases the anticyclonic vortices are
    present  as in the small box case but there is increased activity from
    density waves as seen in the surface density maps.
    These waves could be generated by a process similar to the swing
    amplifier with vorticity source
    described by \cite{HP09a,HP09b}, although the structures
    we observe  do not strictly  correspond to a small scale turbulent flow.
    The density waves are associated with some  outward  angular momentum transport.
    However, the value of $\alpha$ measured from the volume average
    of the Reynolds stress is always highly fluctuating.
    Accordingly we plot running means  as a function of time
   for the small box  and large box 
    with applied viscosity in  Fig. \ref{alphaplot}.
    In the small
    box case there is a small residual time average $\sim 10^{-4}$
    but in the large box this increases to $\sim 3\times 10^{-3}.$
    This is clearly a consequence of the fact that the small box is
    close to  the incompressible regime, whereas the large box
    being effectively larger than a scale height in radial width
    allows the vortices to become large enough to become significantly
    more effective at exciting density waves.

\section{The SBI in 3D\label{3Dincomp}}
The results presented previously were obtained in a 2D setup. It is however known that vortices like the one observed in these simulations are linearly unstable to 3D perturbations due to parametric instabilities \citep{LP09}. The question of the survival of these vortices in 3D is therefore of great importance, as their destruction would lead to the disappearance of the SBI. As shown by \cite{LP09}, 3D instabilities involve relatively small scales compared to the vortex size when the aspect ratio\footnote{The aspect ratio of a vortex is defined by the ratio of the azimuthal size to the radial size of the vorticity patch generated by the vortex} $\chi$ of the vortex is large. This leads to strong numerical constraints on the resolution one has to use to resolve both the 2D SBI and 3D instabilities. To optimise the computational power, we have carried out all the 3D simulations using our spectral code in the Boussinesq approximation, with a setup similar to \S\ref{2Dincomp} and constant stratification.

\subsection{Fiducial simulation\label{3Dnoise}}
\begin{figure}
   \centering
   \includegraphics[width=1.0\linewidth]{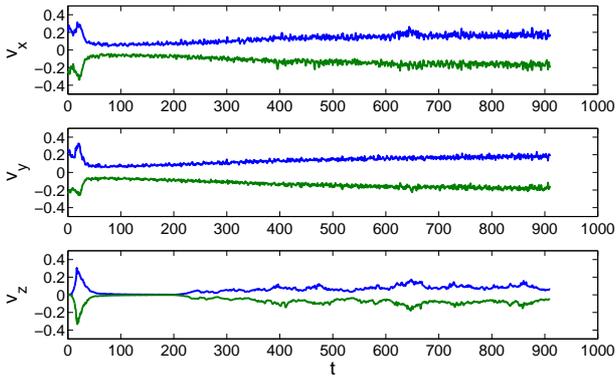}
   \caption{Time history of the maxima of the 3 components of the velocity field when starting from 2D+3D noise. Once the SBI has formed vortices, 3D motions due to parametric instabilities appears and balance the 2D instability.}
              \label{vel3D-noise}%
\end{figure}

\begin{figure*}
   \centering
   \includegraphics[width=0.49\linewidth]{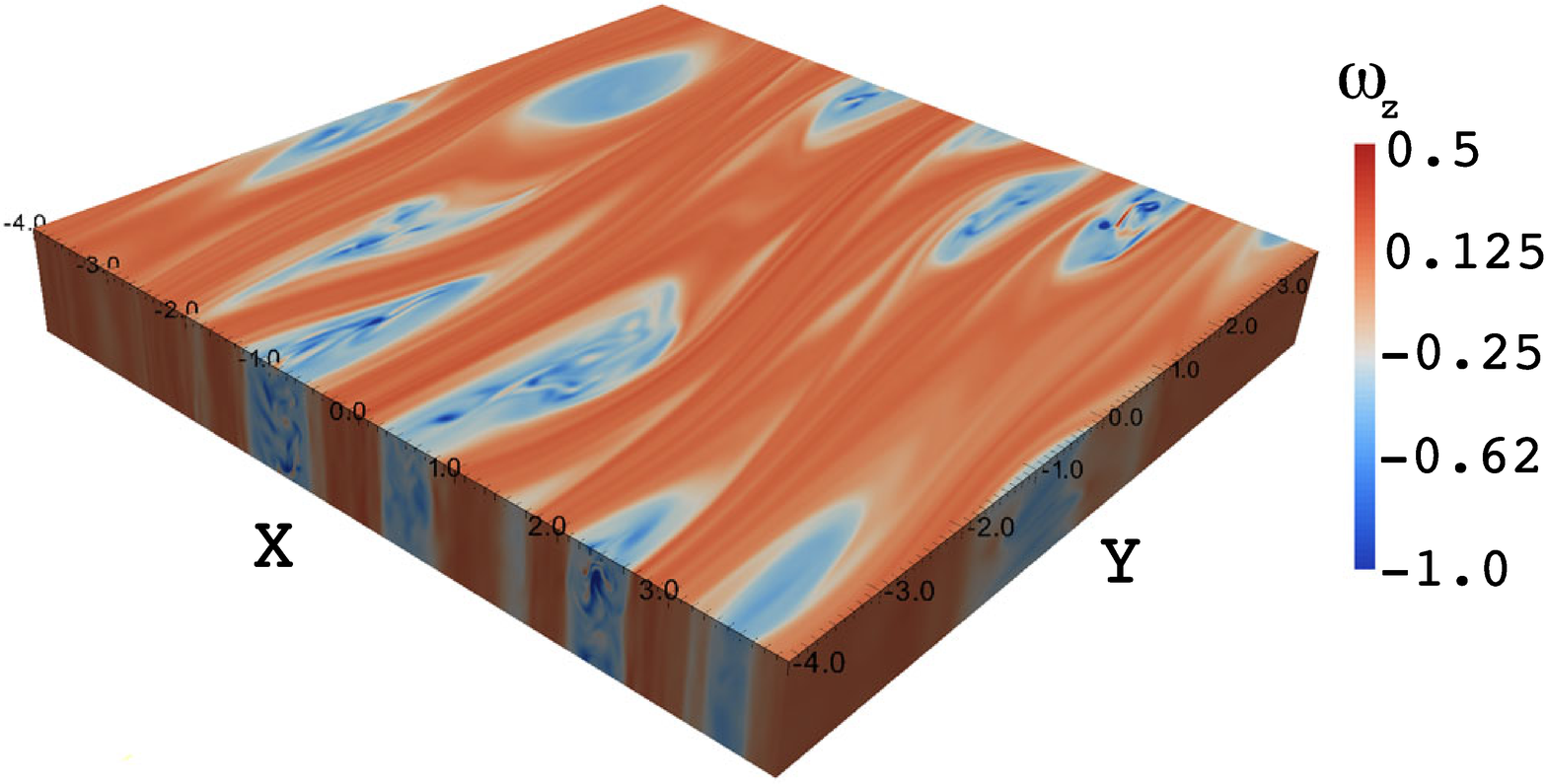}
   \includegraphics[width=0.49\linewidth]{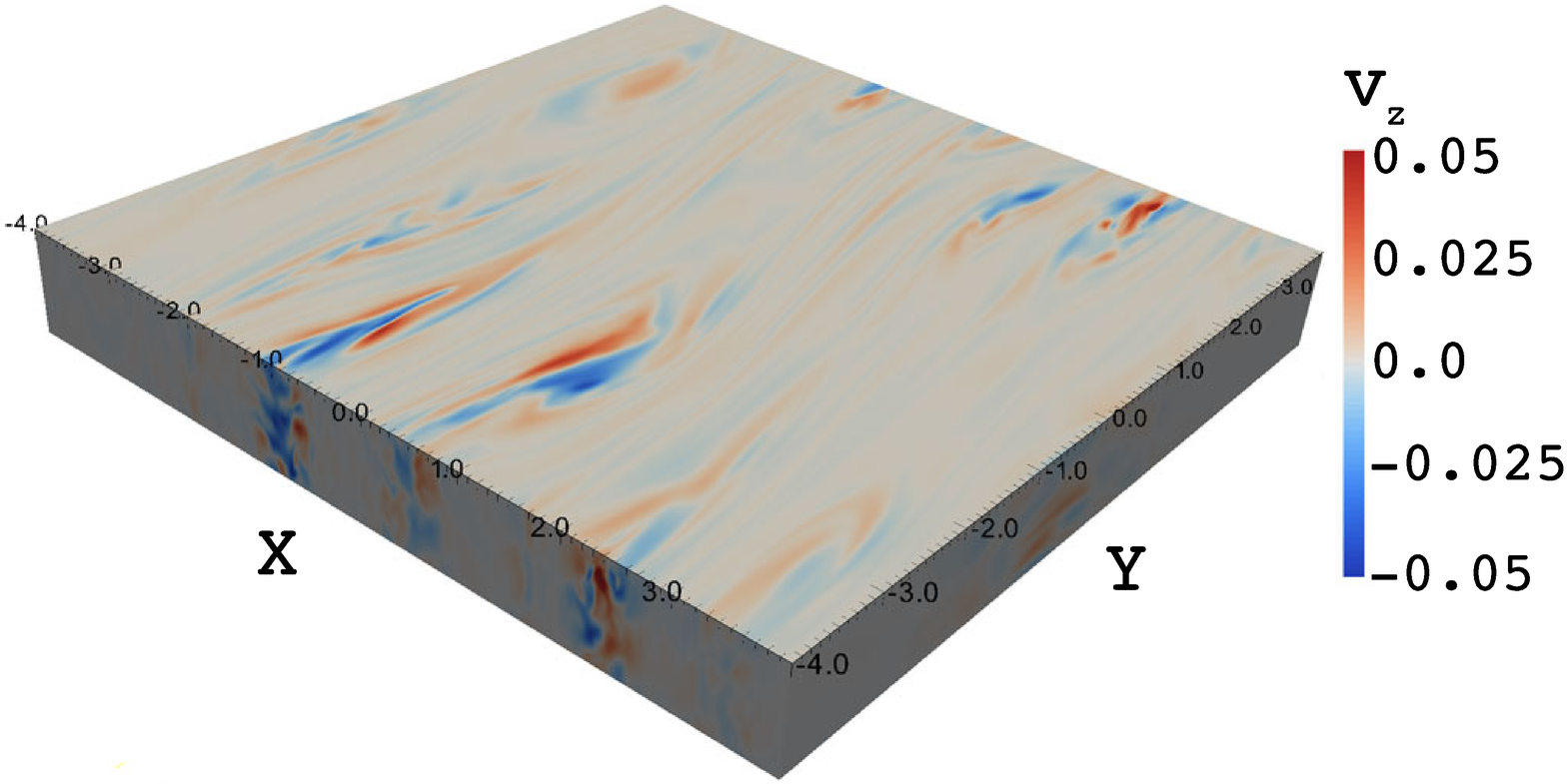}
   
   \caption{3D structure obtained from our fudicial 3D simulation at $t=800$. Left: vertical component of the vorticity. Right: vertical component of the velocity field. We find that 3D instabilities involving vertical motions and vertical structures are localised inside vortex cores.}
              \label{snap3D-noise}%
\end{figure*}

We first consider a box extended horizontally to mimic a thin disc with $L\equiv L_x=L_y=8$ and $L_z=1$. We set $Re=1.02\times 10^6$, $Pe=6400$ and $Ri=-0.01$ with a numerical resolution $N_x\times N_y\times N_z=1024\times 512 \times 128$ in order to resolve the small radial structures due to the 3D instabilities. The horizontal boundary conditions are identical to the one used in 2D and periodic in the vertical direction. Note that we don't include any stratification effect in the vertical direction to reduce computational costs.
Initial conditions are to be chosen with care as 3D random perturbations will generally decay rapidly due to the generation of 3D turbulence everywhere in the box. To avoid this effect, we choose to start with a large amplitude 2D white noise ($\langle \sqrt{v^2}\rangle\sim0.1$) to which we add small 3D perturbations with an amplitude set to 1\% of the 2D perturbation amplitude. 

We show on Fig.\ref{vel3D-noise} the time history of the extrema of the velocity field in such a simulation. As expected, vertical motions triggered by 3D instabilities appear at $t=200$. However, these ``secondary'' instabilities \emph{do not have a destructive effect on the SBI}. Instead, an almost steady state is reached at $t\simeq 600$. 

Looking at the snapshots in the ``saturated state'' at $t=800$ (Fig.~\ref{snap3D-noise}) we observe the production of large scale anticyclonic vortices, as in the 2D case. However, in the core of these vortices (regions of negative vorticity), we also observe the appearance of vertical structures. Looking at the vertical component of the velocity field, one finds clearly that these vertical structures and motions are localised \emph{inside} the vortex core. Although it is likely that the 3D instability observed in this simulation is related to the elliptical instability described by \cite{LP09}, we can't formally prove this point as the vortices found in the simulation are different from the one studied by \cite{LP09}.

Despite the presence of 3D turbulence in these vortices, the turbulent transport measured in this simulation is directed inward, with $\alpha\simeq -3\times 10^{-5}$. This is to be expected as the 3D instabilities extract primarily their energy from the vortex structure and \emph{not} from the mean shear. However, allowing for compressibility will certainly change the turbulent transport as vortices will then also produce density waves (see section \ref{compressibility}).

\subsection{Evolution of a single vortex}
\begin{figure}
   \centering
   \includegraphics[width=1.0\linewidth]{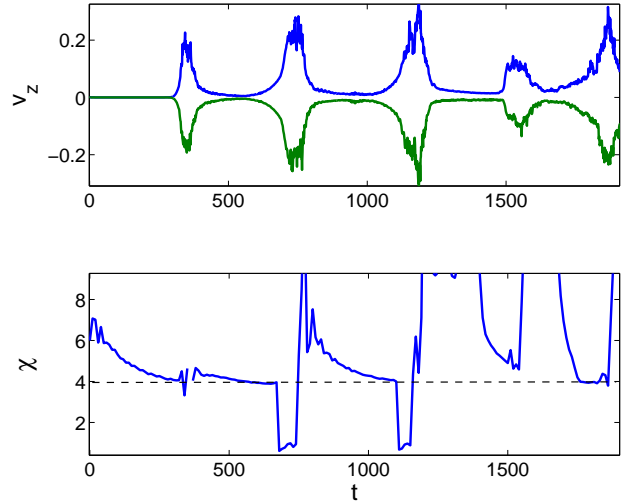}
   \caption{Time history of $v_z$ extrema (top) and of the vortex aspect ratio $\chi$ (bottom). Once the SBI has formed vortices, 3D motions due to parametric instabilities appears and balance the 2D instability.}
              \label{vel3D-kida}%
\end{figure}

\begin{figure*}
   \centering
   \includegraphics[width=0.33\linewidth]{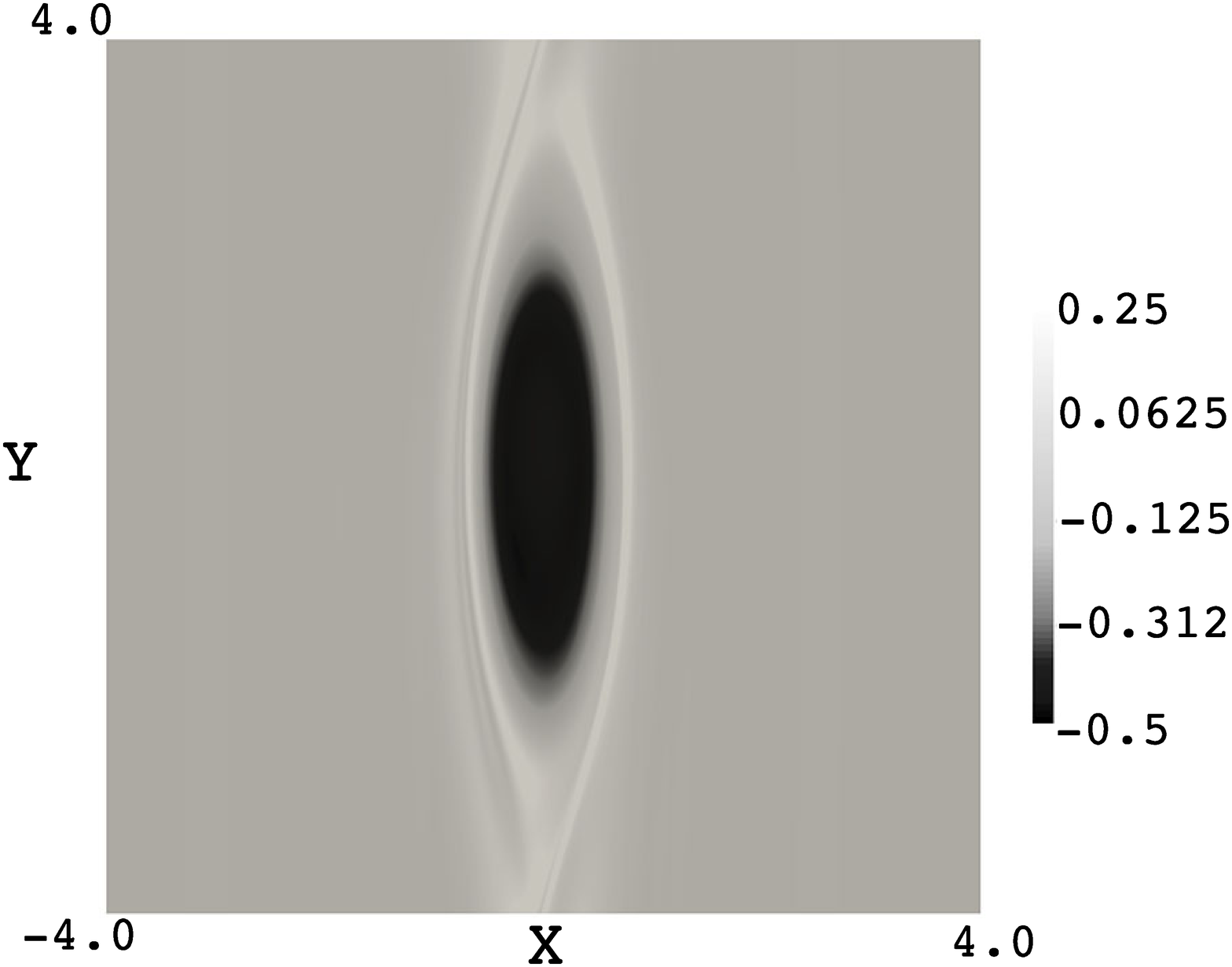}
   \includegraphics[width=0.33\linewidth]{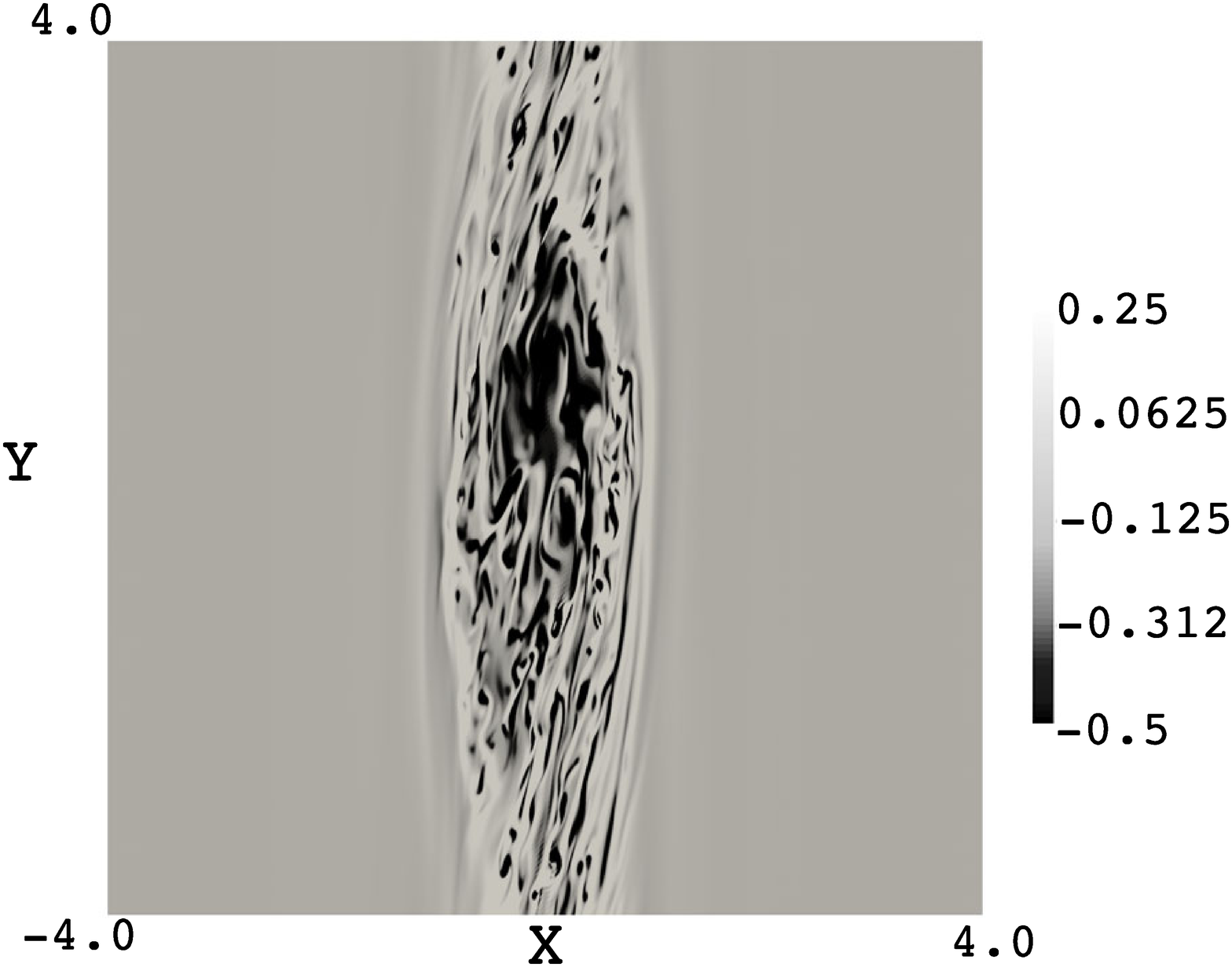}
   \includegraphics[width=0.33\linewidth]{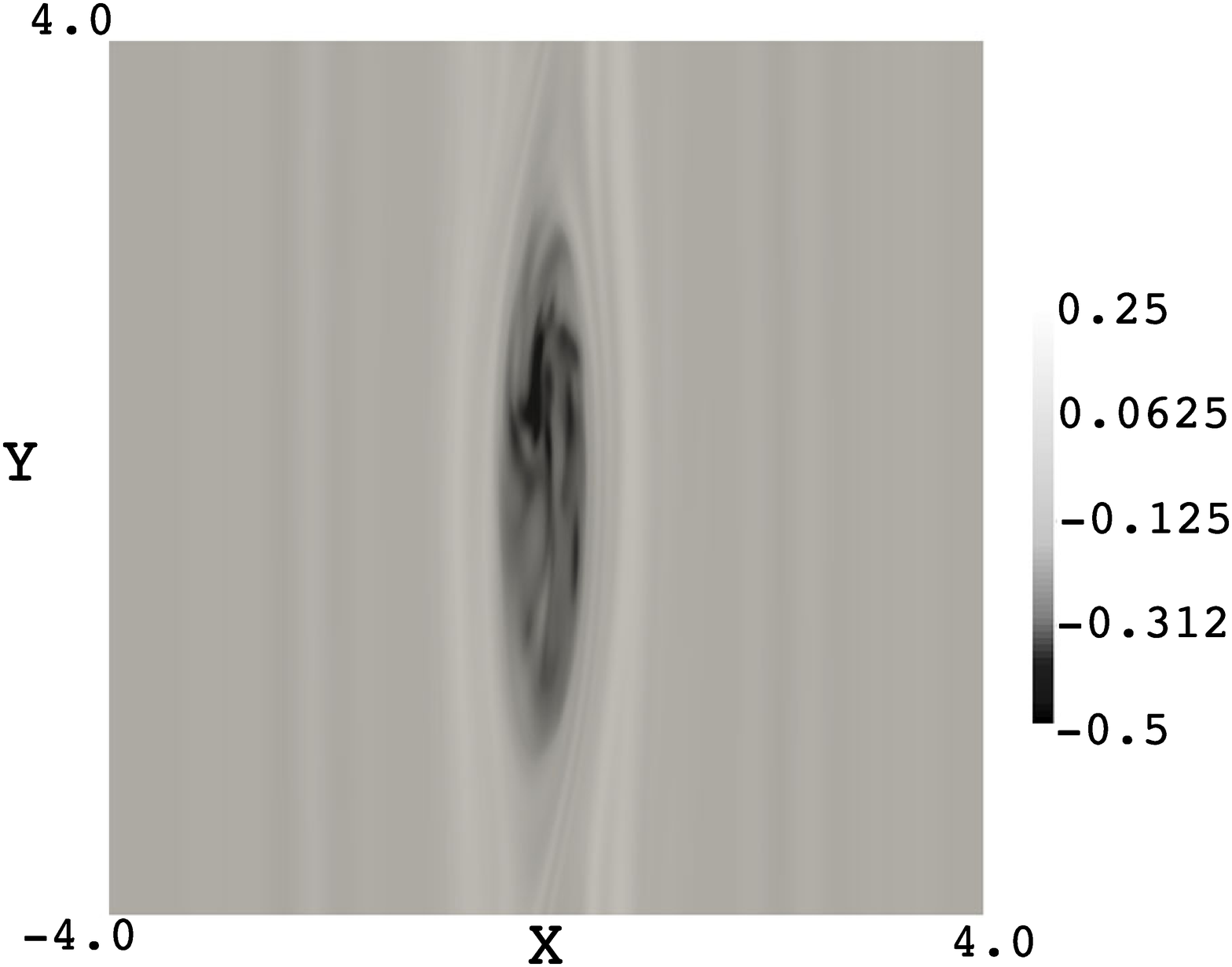}
    
   \caption{Vorticity map from a simulation starting with a 2D Kida vortex plus a weak 3D noise. We show a snapshot at $t=660$ (left), $t=750$ (middle) and $t=830$. After the 3D instability burst, a weak vortex survives and is amplified by the SBI.}
              \label{kida3DSnap}%
\end{figure*}

To isolate the interplay between the SBI and 3D elliptical instabilities, we have considered the case of an isolated 2D vortex in a 3D setup. We take the same parameters as in the fiducial case, but we consider this time a Kida vortex of aspect ratio $\chi=6$ as an initial condition. Thanks to these initial conditions, it is possible to follow the evolution of this single vortex, and in particular measure its aspect ratio as a function of time. This is done with a post-processing script, assuming that the vortex core boundary is located at $\omega_b=0.5[\mathrm{min}(\omega)+\mathrm{max}(\omega)]$. We then measure the vortex aspect ratio as being the ratio $l_y/l_x$ of the boundary defined above. One should note however that this procedure does not check that the vortex is still an ellipse. Moreover, it gives wrong values for $\chi$ if 3D perturbations create strong fluctuations of $\omega$.

We show on Fig.~\ref{vel3D-kida} the result of such a procedure as well as the extrema on the vertical velocity for comparison. Starting from $\chi=6$, the first effect of the SBI is to reduce the vortex aspect ratio. This is consistent with the Kida vortex solution for which
\begin{equation}
\frac{\omega}{S}=-\frac{1}{\chi}\Big(\frac{\chi+1}{\chi-1}\Big).
\end{equation}
Therefore, the vorticity amplification in the vortex core due to the baroclinic feedback leads to a reduction of the aspect ratio. When $\chi\simeq 4$, we note the appearance of strong vertical motions. During these ``bursts'' of 3D turbulence, the aspect ratio measurement is not reliable anymore. However, the vertical motions become ultimately weak, and a weaker vortex appears with $\chi \sim 10$. The baroclinic feedback then amplifies the vortex and the cycle starts again. Interestingly, the $\chi<4$ Kida vortices described by \cite{LP09} were shown to be strongly unstable due to a ``horizontal'' instability. The behaviour observed in these simulations tends to indicate that the parametric modes unstable for $\chi>4$ are not fast enough to take over the SBI. However, as $\chi$ passes through the critical value $\chi=4$, the horizontal modes become unstable and strongly damp the vortex in a few orbits. This interpretation is somewhat confirmed by the snapshots of the velocity field (Fig.~\ref{kida3DSnap}), where growing perturbations are localised inside vortex cores as observed by \cite{LP09}. 

The ``burst'' of turbulence observed in the Kida vortex case might be a specific property of this vortex solution. Indeed, on longer timescales, the spatial distribution of vorticity might be modified, leading to a more progressive appearance of 3D instabilities and a state more similar to the one observed in \ref{3Dnoise} could be achieved. However, this extreme example clearly demonstrate that 3D parametric instabilities do not lead to a total destruction of the vortices produced by the SBI. 

\section{Conclusions\label{conclusions}}
In this paper, we have shown that a subcritical baroclinic instability was found in shearing boxes. Our simulations produce vortices with
a dynamic similar to the description of \cite{PJS07} and \cite{PSJ07}. We find that the conditions required for the SBI are (1) a negative Richardson number (or equivalently a disc unstable
for the radial Schwarzschild criterion) (2) a non negligible thermal diffusion or a cooling function and (3) a finite amplitude initial perturbation, as the instability is subcritical\footnote{Note that point (1) and (2)
were already mentioned by \cite{PSJ07}, although in a somewhat different form.}. 

When computed in compressible shearing boxes, the vortices sustained by the SBI produce density waves which could lead to a \emph{weak} outward angular momentum transport ($\alpha\sim 10^{-3}$), a process suggested by \cite{JG05b}. However, we would like to stress that this transport cannot be approximated by a turbulent viscosity, as the transport due to these waves is certainly a non local process. Several uncertainties remain in the compressible stratified shearing box model.
 In particular, the imposed temperature profile
that is subsequently maintained by a heating source is somewhat artificial.
This occurs while angular momentum transport causes a significant density redistribution causing the local model
to deviate from the original form.
Clearly one  should therefore consider global compressible simulations
with a realistic thermodynamic treatment to draw any firm conclusion on the long term evolution and
 the angular momentum transport aspect of the SBI.

Although the vortices produced by the SBI are anticyclones, the precise pressure structure inside these vortices is not as simple as the description given by \cite{BS95}. In particular, since the vorticity is of the order of the local rotation rate, vortices are not necessarily in geostrophic equilibrium and pressure minima can be found in the centre of strong enough vortices. Moreover, these vortices interact with each other, which leads to complex streamline structures and vortex merging episodes. This leads us to conclude that the particle concentration mechanism proposed by \cite{BS95} might not work in its present form for SBI vortices. 

In 3D, vortices are found to be unstable to parametric instabilities. These instabilities generates random gas motions (``turbulence'') in vortex cores, but they generally don't lead to a proper destruction of the vortex structure itself. More importantly, the presence of a weak hydrodynamic turbulence in vortex cores will lead to a diffusion of dust and boulders, balancing the concentration effect described above. In the end, dust concentration inside these vortices might not be large enough to trigger a gravitational instability and collapse to form planetesimals.

Given the instability criterion detailed above, one can tentatively explain why \cite{JG06} failed to find the SBI in their simulation. First, we would like to stress that our compressible simulations are not more resolved than theirs. Since these simulations use similar numerical schemes, the argument of a too small Reynolds number proposed by \cite{PSJ07} seems dubious. We note however that \cite{JG06} did not include two other important ingredients: a thermal diffusivity and finite amplitude initial perturbations. As shown in this paper, if one of these ingredient is missing, the flow never exhibits the SBI and it gets back to a laminar state rapidly. Note that even if this argument explains the negative result of \cite{JG06}, it also indicates that the SBI should be absent from the simulations presented by \cite{KB03}, as no thermal diffusion nor cooling function was used 
(the initial conditions were not explicitly mentioned). This suggests that either the global baroclinic instability described by \cite{KB03} and the SBI are two separate instabilities, or strong numerical artefacts have produced an artificially large thermal diffusion in \cite{KB03} simulations. 

In the limit of a very small kinematic viscosity, the scale at which the instability appears has to be related to the diffusion length scale $(\mu/S)^{1/2}$. In a disc, one would therefore expect the instability around that scale, provided that the entropy profile has the right slope (condition 1). However, as the instability produces enstrophy, vortices are expected to grow, until they reach a size of the order of the disc thickness where compressible effects balance the SBI source term. We conclude from this argument that even if the SBI itself is found at small scale (e.g. because of a small thermal diffusivity), the end products of this instability are large scale vortices, with a size of the order of a few disc scale heights.

We would like to stress that the present work constitutes only a \emph{proof of concept} for the subcritical baroclinic instability. To claim that this instability actually exists in discs, one has to study the disc thermodynamic in a self consistent manner, a task which is well beyond the scope of this paper. Moreover, it is not possible at this stage to state that the SBI is a solution to the angular momentum transport problem in discs. Although this instability creates some transport through the generation of waves, this transport is weak and large uncertainties remain on its exact value. Finally, the coexistence of the SBI and the MRI \citep{BH91a} is for the moment unclear. In the presence of magnetic fields, vortices produced by the SBI might become strongly unstable because of magnetoelliptic instabilities \citep{MB09}. Although the nonlinear outcome of these instabilities is unknown, one can already suspect that vortices produced by the SBI will be weaken in the presence of magnetic fields.

\begin{acknowledgements}
      GL thanks Charles Gammie, Hubert Klahr, Pierre-Yves Longaretti and Wladimir Lyra for useful and stimulating discussions during the Newton Institute program on the dynamics of discs and planets. We thank our referee G. Stewart for his valuable suggestions on this work. The simulations presented in this paper were performed using the Darwin Supercomputer of the University of Cambridge High Performance Computing Service (http://www.hpc.cam.ac.uk/), provided by Dell Inc. using Strategic Research Infrastructure Funding from the Higher Education Funding Council for England. GL acknowledges support by STFC . 
\end{acknowledgements}

\bibliographystyle{aa}
\bibliography{glesur}

\begin{thebibliography}{37}
\expandafter\ifx\csname natexlab\endcsname\relax\def\natexlab#1{#1}\fi

\bibitem[{{Balbus}(2003)}]{B03}
{Balbus}, S.~A. 2003, \araa, 41, 555

\bibitem[{{Balbus} \& {Hawley}(1991)}]{BH91a}
{Balbus}, S.~A. \& {Hawley}, J.~F. 1991, \apj, 376, 214

\bibitem[{{Barge} \& {Sommeria}(1995)}]{BS95}
{Barge}, P. \& {Sommeria}, J. 1995, \aap, 295, L1

\bibitem[{{Barranco} \& {Marcus}(2005)}]{BM05}
{Barranco}, J.~A. \& {Marcus}, P.~S. 2005, \apj, 623, 1157

\bibitem[{{Chagelishvili} {et~al.}(2003){Chagelishvili}, {Zahn}, {Tevzadze}, \&
  {Lominadze}}]{CZTL03}
{Chagelishvili}, G.~D., {Zahn}, J.-P., {Tevzadze}, A.~G., \& {Lominadze}, J.~G.
  2003, \aap, 402, 401

\bibitem[{{de Val-Borro} {et~al.}(2007){de Val-Borro}, {Artymowicz},
  {D'Angelo}, \& {Peplinski}}]{VAAP07}
{de Val-Borro}, M., {Artymowicz}, P., {D'Angelo}, G., \& {Peplinski}, A. 2007,
  \aap, 471, 1043

\bibitem[{{de Val-Borro} {et~al.}(2006){de Val-Borro}, {Edgar}, {Artymowicz},
  {Ciecielag}, {Cresswell}, {D'Angelo}, {Delgado-Donate}, {Dirksen}, {Fromang},
  {Gawryszczak}, {Klahr}, {Kley}, {Lyra}, {Masset}, {Mellema}, {Nelson},
  {Paardekooper}, {Peplinski}, {Pierens}, {Plewa}, {Rice}, {Sch{\"a}fer}, \&
  {Speith}}]{VE06}
{de Val-Borro}, M., {Edgar}, R.~G., {Artymowicz}, P., {et~al.} 2006, \mnras,
  370, 529

\bibitem[{{Fricke}(1968)}]{F68}
{Fricke}, K. 1968, Zeitschrift fur Astrophysik, 68, 317

\bibitem[{{Fromang} \& {Nelson}(2005)}]{FN05}
{Fromang}, S. \& {Nelson}, R.~P. 2005, \mnras, 364, L81

\bibitem[{{Fromang} \& {Papaloizou}(2006)}]{FP06}
{Fromang}, S. \& {Papaloizou}, J. 2006, \aap, 452, 751

\bibitem[{{Fromang} {et~al.}(2007){Fromang}, {Papaloizou}, {Lesur}, \&
  {Heinemann}}]{FPLH07}
{Fromang}, S., {Papaloizou}, J., {Lesur}, G., \& {Heinemann}, T. 2007, \aap,
  476, 1123

\bibitem[{{Goldreich} \& {Schubert}(1967)}]{GS67}
{Goldreich}, P. \& {Schubert}, G. 1967, \apj, 150, 571

\bibitem[{{Hawley} {et~al.}(1999){Hawley}, {Balbus}, \& {Winters}}]{HBW99}
{Hawley}, J.~F., {Balbus}, S.~A., \& {Winters}, W.~F. 1999, \apj, 518, 394

\bibitem[{{Hawley} {et~al.}(1995){Hawley}, {Gammie}, \& {Balbus}}]{HGB95}
{Hawley}, J.~F., {Gammie}, C.~F., \& {Balbus}, S.~A. 1995, \apj, 440, 742

\bibitem[{{Heinemann} \& {Papaloizou}(2009{\natexlab{a}})}]{HP09a}
{Heinemann}, T. \& {Papaloizou}, J.~C.~B. 2009{\natexlab{a}}, \mnras, 397, 52

\bibitem[{{Heinemann} \& {Papaloizou}(2009{\natexlab{b}})}]{HP09b}
{Heinemann}, T. \& {Papaloizou}, J.~C.~B. 2009{\natexlab{b}}, \mnras, 397, 64

\bibitem[{{Ji} {et~al.}(2006){Ji}, {Burin}, {Schartman}, \& {Goodman}}]{JBSG06}
{Ji}, H., {Burin}, M., {Schartman}, E., \& {Goodman}, J. 2006, \nat, 444, 343

\bibitem[{{Johnson} \& {Gammie}(2005{\natexlab{a}})}]{JG05}
{Johnson}, B.~M. \& {Gammie}, C.~F. 2005{\natexlab{a}}, \apj, 626, 978

\bibitem[{{Johnson} \& {Gammie}(2005{\natexlab{b}})}]{JG05b}
{Johnson}, B.~M. \& {Gammie}, C.~F. 2005{\natexlab{b}}, \apj, 635, 149

\bibitem[{{Johnson} \& {Gammie}(2006)}]{JG06}
{Johnson}, B.~M. \& {Gammie}, C.~F. 2006, \apj, 636, 63

\bibitem[{{Kida}(1981)}]{K81}
{Kida}, S. 1981, Physical Society of Japan, Journal, vol.~50, Oct.~1981,
  p.~3517-3520., 50, 3517

\bibitem[{{Klahr}(2004)}]{K04}
{Klahr}, H. 2004, \apj, 606, 1070

\bibitem[{{Klahr} \& {Bodenheimer}(2003)}]{KB03}
{Klahr}, H.~H. \& {Bodenheimer}, P. 2003, \apj, 582, 869

\bibitem[{{Kley} {et~al.}(1993){Kley}, {Papaloizou}, \& {Lin}}]{KPL93}
{Kley}, W., {Papaloizou}, J.~C.~B., \& {Lin}, D.~N.~C. 1993, \apj, 409, 739

\bibitem[{{Lesur} \& {Longaretti}(2005)}]{LL05}
{Lesur}, G. \& {Longaretti}, P.-Y. 2005, \aap, 444, 25

\bibitem[{{Lesur} \& {Longaretti}(2007)}]{LL07}
{Lesur}, G. \& {Longaretti}, P.-Y. 2007, \mnras, 378, 1471

\bibitem[{{Lesur} \& {Papaloizou}(2009)}]{LP09}
{Lesur}, G. \& {Papaloizou}, J.~C.~B. 2009, \aap, 498, 1

\bibitem[{{Lovelace} {et~al.}(1999){Lovelace}, {Li}, {Colgate}, \&
  {Nelson}}]{LLC99}
{Lovelace}, R.~V.~E., {Li}, H., {Colgate}, S.~A., \& {Nelson}, A.~F. 1999,
  \apj, 513, 805

\bibitem[{{Mizerski} \& {Bajer}(2009)}]{MB09}
{Mizerski}, K.~A. \& {Bajer}, K. 2009, Journal of Fluid Mechanics, 632, 401

\bibitem[{{Papaloizou} {et~al.}(2004){Papaloizou}, {Nelson}, \&
  {Snellgrove}}]{PNS04}
{Papaloizou}, J.~C.~B., {Nelson}, R.~P., \& {Snellgrove}, M.~D. 2004, \mnras,
  350, 829

\bibitem[{{Petersen} {et~al.}(2007{\natexlab{a}}){Petersen}, {Julien}, \&
  {Stewart}}]{PJS07}
{Petersen}, M.~R., {Julien}, K., \& {Stewart}, G.~R. 2007{\natexlab{a}}, \apj,
  658, 1236

\bibitem[{{Petersen} {et~al.}(2007{\natexlab{b}}){Petersen}, {Stewart}, \&
  {Julien}}]{PSJ07}
{Petersen}, M.~R., {Stewart}, G.~R., \& {Julien}, K. 2007{\natexlab{b}}, \apj,
  658, 1252

\bibitem[{{Regev} \& {Umurhan}(2008)}]{RU08}
{Regev}, O. \& {Umurhan}, O.~M. 2008, \aap, 481, 21

\bibitem[{{Spiegel} \& {Veronis}(1960)}]{SV60}
{Spiegel}, E.~A. \& {Veronis}, G. 1960, \apj, 131, 442

\bibitem[{{Umurhan} \& {Regev}(2004)}]{UR04}
{Umurhan}, O.~M. \& {Regev}, O. 2004, \aap, 427, 855

\bibitem[{{von Weizs\"acker}(1944)}]{W44}
{von Weizs\"acker}, C.~F. 1944, Zeit. f\"ur Astrophys., 22, 319

\bibitem[{{Ziegler} \& {Yorke}(1997)}]{ZY97}
{Ziegler}, U. \& {Yorke}, H.~W. 1997, Computer Physics Communications, 101, 54

\end{thebibliography}

\end{document}